\documentclass[aps,prb,twocolumn,showpacs,floatfix]{revtex4-1}
\usepackage{graphicx,amsfonts,amssymb,amsmath,hyperref}
\usepackage{textcomp}
\usepackage{esint}

\newif\ifhyper
\hypertrue
\ifhyper       
\hypersetup{        
  citecolor = {green},
  colorlinks = {true}, 
  urlcolor = {blue} 
}

\begin{document}

\def\Gch{G_{\rm ch}} 
\def\Gsp{G_{\rm sp}} 
\def\Gchk{G_{k,\rm ch}} 
\def\Gspk{G_{k,\rm sp}} 
\def\Gspv{{\bf G}_{\rm sp}} 
\def\Gspvk{{\bf G}_{k,\rm sp}} 
\def\bGch{\bar G_{\rm ch}} 
\def\bGsp{\bar G_{\rm sp}} 
\def\bGchk{\bar G_{k,\rm ch}} 
\def\bGspk{\bar G_{k,\rm sp}} 
\def\bGspv{\bar {\bf G}_{\rm sp}} 
\def\bGspvk{\bar {\bf G}_{k,\rm sp}} 
\def\Fs{F_{\rm s}}
\def\Ft{F_{\rm t}}
\def\Fsk{F_{k,\rm s}}
\def\Ftk{F_{k,\rm t}}
\def\Ftv{\F_{\rm t}}
\def\Ftvk{\F_{k,\rm t}}
\def\bFs{\bar F_{\rm s}}
\def\bFt{\bar F_{\rm t}}
\def\bFsk{\bar F_{k,\rm s}}
\def\bFtk{\bar F_{k,\rm t}}
\def\bFtv{\bar \F_{\rm t}}
\def\bFtvk{\bar \F_{k,\rm t}}
\def\Uch{U_{\rm ch}}
\def\Usp{U_{\rm sp}}
\def\Us{U_{\rm s}}
\def\Ut{U_{\rm t}}
\def\uch{u_{\rm ch}}
\def\usp{u_{\rm sp}}
\def\us{u_{\rm s}}
\def\ut{u_{\rm t}}

\def\Gkch{G_{k,\rm ch}} 
\def\Gksp{G_{k,\rm sp}} 
\def\Gkspv{{\bf G}_{k,\rm sp}} 
\def\bGkch{\bar G_{k,\rm ch}} 
\def\bGksp{\bar G_{k,\rm sp}} 
\def\bGkspv{\bar {\bf G}_{k,\rm sp}} 
\def\Fks{F_{k,\rm s}}
\def\Fkt{F_{k,\rm t}}
\def\Fktv{\F_{k,\rm t}}
\def\bFks{\bar F_{k,\rm s}}
\def\bFkt{\bar F_{k,\rm t}}
\def\bFktv{\bar \F_{k,\rm t}}
\def\ukch{u_{k,\rm ch}}
\def\uksp{u_{k,\rm sp}}
\def\uks{u_{k,\rm s}}
\def\ukt{u_{k,\rm t}}


\graphicspath{{./figures_submit/}}
\def\rhoeq{\hat\rho_{\rm eq}}

\newcommand{\marge}[1]{\marginpar{\scriptsize #1}}
\newcommand{\remarque}[1]{\marginpar{\scriptsize Remarque}{\it [#1]}}
\newcommand{\new}[1]{{\bf #1}}

\def\beq{\begin{equation}}
\def\eeq{\end{equation}}
\def\bleq{\begin{eqnarray}}
\def\eleq{\end{eqnarray}} 
\def\bfig{\begin{figure}}
\def\efig{\end{figure}}
\def\bline{\begin{multline}}
\def\eline{\end{multline}}
\def\bremark{\begin{quotation} \noindent \small }
\def\eremark{\end{quotation}}
\def\llbrace{\left\lbrace}
\def\rrbrace{\right\rbrace}
\def\lbraket{\left[}
\def\rbraket{\right]}
\def\llangle{\left\langle}
\def\rrangle{\right\rangle} 

\newcommand{\Tr}{{\rm Tr}} 
\newcommand{\tr}{{\rm tr}} 
\newcommand{\sgn}{{\rm sgn}} 
\newcommand{\mean}[1]{\langle #1 \rangle}
\newcommand{\commu}[2]{[#1,#2]} 
\newcommand{\bra}[1]{\langle#1|}
\newcommand{\ket}[1]{|#1\rangle}
\newcommand{\braket}[2]{\langle #1|#2\rangle}
\newcommand{\dbraket}[3]{\langle #1|#2|#3\rangle}
\newcommand{\tens}[1]{\overleftrightarrow{#1}}  
\newcommand{\vac}{|{\rm vac}\rangle} 
\def\bravac{\langle{\rm vac}|}
\newcommand{\const}{{\rm const}} 
\newcommand{\atanh}{\,{\rm atanh}}

\newcommand{\ie}{i.e. }
\newcommand{\iet}{i.e.}
\newcommand{\eg}{e.g. }
\newcommand{\cc}{{\rm c.c.}} 
\newcommand{\hc}{{\rm h.c.}} 
\def\etal{{\it et al. }}

\newcommand{\jhatbf}{\hat {\textbf \j}} 
\newcommand{\Jhatbf}{\hat {\textbf \J}} 
\newcommand{\jhat}{\hat {\jmath}} 
\newcommand{\Jhat}{\hat {J}} 
\newcommand{\jbf}{\textbf j}
\newcommand{\Jbf}{\textbf J}

\def\chibf{\boldsymbol{\chi}}
\def\down{\downarrow}
\def\eps{\epsilon}
\def\gam{\gamma} 
\def\phibf{\boldsymbol{\phi}}
\def\varphibf{\boldsymbol{\varphi}}
\def\varphibfs{\boldsymbol{\varphi}_<}
\def\varphibfl{\boldsymbol{\varphi}_>}
\def\varphis{\varphi_{<}}
\def\varphil{\varphi_{>}}
\def\psibf{\boldsymbol{\psi}}
\def\Ome{\Omega}
\def\omeD{{\omega_D}} 
\def\bfOme{\boldsymbol{\Omega}} 
\def\Omebf{\boldsymbol{\Omega}} 
\def\lamb{\lambda}
\def\Lamb{\Lambda}
\def\sig{\sigma}
\def\Sig{\Sigma}
\def\sigp{{\sigma'}} 
\def\bfsig{\boldsymbol{\sigma}} 
\def\sigbf{\boldsymbol{\sigma}} 
\def\The{\Theta} 
\def\up{\uparrow}

\def\epsk{\epsilon_{\bf k}} 
\def\xik{\xi_{\bf k}} 
\def\xip{\xi_{\bf p}} 
\def\xikq{\xi_{{\bf k}+{\bf q}}} 
\def\Ek{E_{\bf k}} 
\def\Ep{E_{\bf p}}
\def\Heff{\hat H_{\rm eff}}
\def\Hem{\hat H_{\rm em}}
\def\Hint{\hat H_{\rm int}}
\def\Hloc{\hat H_{\rm loc}}
\def\HMF{\hat H_{\rm MF}}
\def\Sem{S_{\rm em}}
\def\SMF{S_{\rm MF}} 
\def\SRPA{S_{\rm RPA}} 
\def\Sint{S_{\rm int}} 
\def\Sloc{S_{\rm loc}} 
\def\Zloc{Z_{\rm loc}} 
\def\ZMF{Z_{\rm MF}} 
\def\ZRPA{Z_{\rm RPA}} 
\def\RPA{{\rm RPA}}
\def\loc{{\rm loc}} 
\def\pp{{\rm pp}}
\def\ph{{\rm ph}} 
\def\ch{{\rm ch}}
\def\sp{{\rm sp}} 
\def\qtf{q_{\rm TF}}
\def\epstf{\eps^{}_{\rm TF}} 
\def\epsrpa{\eps^{}_{\rm RPA}} 
\def\chinnzpp{\chi_{nn}^{0}{}\!\!\!''}

\def\half{\frac{1}{2}}
\def\dhalf{\dfrac{1}{2}}
\def\third{\frac{1}{3}} 
\def\quarter{\frac{1}{4}}

\def\qr{{\bf q}\cdot{\bf r}}
\def\wt{\omega t} 

\def\a{{\bf a}}
\def\b{{\bf b}}
\def\e{{\bf e}}
\def\f{{\bf f}}
\def\g{{\bf g}}
\def\h{{\bf h}}
\def\k{{\bf k}}
\def\l{{\bf l}}
\def\m{{\bf m}}
\def\n{{\bf n}} 
\def\p{{\bf p}} 
\def\q{{\bf q}}
\def\r{{\bf r}}
\def\t{{\bf t}}
\def\u{{\bf u}}
\def\v{{\bf v}}
\def\x{{\bf x}}
\def\y{{\bf y}} 
\def\z{{\bf z}} 
\def\A{{\bf A}}
\def\B{{\bf B}}
\def\D{{\bf D}} 
\def\E{{\bf E}} 
\def\F{{\bf F}} 
\def\H{{\bf H}}  
\def\J{{\bf J}}
\def\K{{\bf K}} 

\def\G{{\bf G}}
\def\L{{\bf L}}
\def\M{{\bf M}}  
\def\O{{\bf O}} 
\def\P{{\bf P}} 
\def\Q{{\bf Q}} 
\def\R{{\bf R}}
\def\S{{\bf S}}
\def\epsbf{\boldsymbol{\epsilon}}
\def\mubf{\boldsymbol{\mu}}
\def\nablabf{\boldsymbol{\nabla}}
\def\rhobf{\boldsymbol{\rho}}
\def\sigmabf{\boldsymbol{\sigma}} 
\def\Pibf{\boldsymbol{\Pi}}
\def\pibf{\boldsymbol{\pi}}

\def\para{\parallel}
\def\kpara{{k_\parallel}}
\def\kperp{{k_\perp}} 
\def\kperpp{{k_\perp'}} 
\def\qperp{{q_\perp}} 
\def\tperp{{t_\perp}} 

\def\w{\omega}
\def\wn{\omega_n}
\def\wnu{\omega_\nu}
\def\wp{\omega_p} 
\def\dmu{{\partial_\mu}}
\def\dl{{\partial_l}}  
\def\dt{\partial_t} 
\def\tdt{\tilde\partial_t}
\def\dk{\partial_k}
\def\tdk{\tilde\partial_k}
\def\dx{\partial_x}
\def\dy{\partial_y} 
\def\dtau{{\partial_\tau}}  
\def\det{{\rm det}} 
\def\Pf{{\rm Pf}}

\def\dsum{\displaystyle \sum}
\def\dint{\displaystyle \int} 
\def\intt{\int_{-\infty}^\infty dt} 
\def\inttp{\int_{-\infty}^\infty dt'} 
\def\intk{\int_{\bf k}} 
\def\intkd{\int \frac{d^dk}{(2\pi)^d}}
\def\intq{\int_{\bf q}} 
\def\intr{\int d^dr}  
\def\dintr{\displaystyle \int d^dr} 
\def\intrp{\int d^dr'}
\def\dinttau{\displaystyle \int_0^\beta d\tau}
\def\dinttaup{\displaystyle \int_0^\beta d\tau'}
\def\inttau{\int_0^\beta d\tau}
\def\inttaup{\int_0^\beta d\tau'}
\def\intx{\int d^{d+1}x} 
\def\inttaur{\int_0^\beta d\tau \int d^dr}
\def\intinf{\int_{-\infty}^\infty}
\def\dinttaur{\displaystyle \int_0^\beta d\tau \int d^dr}
\def\dintinf{\displaystyle \int_{-\infty}^\infty}
\def\intw{\int_{-\infty}^\infty \frac{d\w}{2\pi}}
\def\sumr{\sum_{\bf r}} 

\def\calA{{\cal A}} 
\def\calC{{\cal C}} 
\def\dt{\partial_t}
\def\calD{{\cal D}}
\def\calF{{\cal F}} 
\def\calG{{\cal G}}
\def\calH{{\cal H}}
\def\calI{{\cal I}}
\def\calJ{{\cal J}}
\def\calK{{\cal K}}
\def\calL{{\cal L}} 
\def\calN{{\cal N}}
\def\calO{{\cal O}}
\def\calP{{\cal P}}  
\def\calR{{\cal R}} 
\def\calS{{\cal S}}
\def\calT{{\cal T}}
\def\calU{{\cal U}}
\def\calX{{\cal X}} 
\def\calY{{\cal Y}} 
\def\calZ{{\cal Z}} 

\def\calFbf{{\bf F}}

\def\tT{{\tilde T}}
\def\talpha{{\tilde\alpha}}
\def\tdelta{{\tilde\delta}}
\def\teta{{\tilde\eta}} 
\def\tlamb{{\tilde\lambda}}
\def\tmu{{\tilde\mu}}
\def\tphibf{{\tilde\phibf}}
\def\trho{{\tilde\rho}}
\def\tvarphibf{{\tilde\varphibf}} 
\def\tw{{\tilde\omega}}
\def\twn{{\tilde\omega_n}}

\def\asinh{{\rm asinh}} 

\title{Nonperturbative renormalization-group approach to fermion systems in the two-particle-irreducible effective action formalism} 

\author{N. Dupuis}
\affiliation{Laboratoire de Physique Th\'eorique de la Mati\`ere Condens\'ee, 
CNRS UMR 7600, Universit\'e Pierre et Marie Curie, 4 Place Jussieu, 
75252 Paris Cedex 05, France}

\date{December 13, 2013} 

\begin{abstract}
We propose a nonperturbative renormalization-group (NPRG) approach to fermion systems in the two-particle-irreducible (2PI) effective action formalism, based on an exact RG equation for the Luttinger-Ward functional. This approach enables us to describe phases with spontaneously broken symmetries while satisfying the Mermin-Wagner theorem. We show that it is possible to choose the Hartree-Fock--RPA theory as initial condition of the RG flow and argue that the 2PI-NPRG is not restricted to the weak-coupling limit. An expansion of the Luttinger-Ward functional about the minimum of the 2PI effective action including only the two-particle 2PI vertex leads to nontrivial RG equations where interactions between fermions and collective excitations naturally emerge.   
\end{abstract}
\maketitle

\section{Introduction}

The renormalization group (RG) has proven to be a systematic and unbiased method to study interacting fermion systems. In contrast to standard (ladder- or bubble-type) diagrammatic resummations, it treats various types of instabilities on an equal footing and does not require any {\it a priori} knowledge of the ground state of the system. The RG has been particularly useful to understand one- and quasi-one-dimensional systems~\cite{Solyom79,Bourbonnais91} in connection with the physics of strongly anisotropic organic conductors (see, e.g., Refs.~\onlinecite{Bourbonnais88,Duprat01,Nickel05,Nickel06,Bourbonnais09,Sedeki12}). It has also been used to study the two-dimensional Hubbard model (see, e.g., Refs.~\onlinecite{Zanchi00,Halboth00,Honerkamp01,Husemann09,Husemann12,Giering12}) as well as other models of strongly correlated fermions (for a review, see Ref.~\onlinecite{Metzner12}).

Nevertheless the fermionic RG meets with some difficulties: i) it is usually formulated in the one-particle-irreducible (1PI) formalism where the basic quantity of interest is the 1PI effective action $\Gamma[\phi^*,\phi]$ (the generating functional of 1PI vertices). While for a classical system or bosons $\phi$ is a real or complex field, for fermions it is a Grassmannian (anticommuting) field. The effective action is therefore defined only {\it via} its Taylor expansion about $\phi^*=\phi=0$. Such an expansion, truncated to a given order, is equivalent to a loop expansion. Expansions about a nontrivial minimum, which are very efficient in the standard implementation of the nonperturbative renormalization group (NPRG) even with low-order truncations,\cite{Berges02,Delamotte12,Kopietz_book} are not possible with fermions. This makes calculations beyond  one- or two-loop order difficult and restricts the fermionic RG to the weak-coupling limit. ii) The existence of a Fermi surface implies that the interaction 
amplitudes strongly depend on the fermion momenta, 
so that the renormalized interaction vertices necessary become functionals of momenta.\cite{Shankar94,note11} This considerably increases the complexity of the method, even to one-loop order. iii) The fermion field $\psi$ is not an order parameter. Order parameters are defined by composite fields: $\psi^*_\sig\psi_{\sig'}$ for a charge- or spin-density wave, $\psi_\sig\psi_{\sig'}$ for a superconductor ($\psi_\sig$ denotes the fermion field and $\sig$ the spin index). Phase transitions are signaled by a divergence of the order parameter susceptibility and a concomitant divergence of some interaction vertices. In the standard implementation of the fermionic RG, these composite fields are not explicitly considered since the RG procedure deals only with the fermionic degrees of freedom.\cite{note10} This explains why some vertices or susceptibilities may diverge at a finite energy scale with no possibility to continue the flow into the broken-symmetry phase.~\cite{note15} Even in the absence of a phase 
transition, the 
one-loop fermionic RG becomes uncontrolled when strong collective fluctuations with a large correlation length set in. 

To circumvent some of these difficulties, one can partially ``bosonize'' the fermionic degrees of freedom {\it via} a Hubbard-Stratonovich transformation of the interaction term in the action $S[\psi^*,\psi]$.\cite{Baier00,Baier04,Baier05,Schutz05,Gies02} This transforms the original interacting fermion system into a system of free fermions interacting with a bosonic field. While the difficulty to consider various types of instabilities on an equal footing is well known (see, however, Refs.~\onlinecite{Krahl07,Krahl09a,Friederich11}), it is possible to treat the bosonic degrees of freedom with the standard methods of the NPRG. There are then no difficulties to study phases with spontaneously broken symmetries since the bosonic Hubbard-Stratonovich field plays the role of an order parameter. Furthermore the approach is not restricted to the weak-coupling limit as it partially relies on the NPRG. Both the Hubbard model\cite{Baier00,Baier04,Baier05,Krahl07,Krahl09a,Friederich11} and the BCS-BEC crossover\cite{
Birse05,
Diehl07a,Diehl07b,Floerchinger08a,Scherer10,Strack08,Bartosch09,Eberlein13a} have been studied along these lines.

The main purpose of this paper is to discuss a NPRG approach in the two-particle-irreducible (2PI) formalism following general ideas put forward by Wetterich.\cite{Wetterich07} We restrict our aim to a discussion of the general aspects of the method and postpone practical applications to future work. In condensed-matter physics, the 2PI formalism\cite{Luttinger60b,Baym61,Baym62,Dedominicis64a,Dedominicis64b,[{The 2PI formalism has been extended to relativistic field theories by }]Cornwall74} was introduced as a means to systematically set up self-consistent approximations that satisfy conservation laws (conserving approximations\cite{Baym62}). The basic quantity of interest is the 2PI effective action $\Gamma[\calG]$ or equivalently the Luttinger-Ward functional $\Phi[\calG]$, a functional of the one-particle propagator $\calG$ defined perturbatively as the sum of the 2PI (or skeleton) diagrams. $\Phi[\calG]$ is the generating functional of the 2PI vertices from which we can obtain both 1PI vertices and 
correlation functions. The 2PI formalism in the context of the NPRG has been considered in various contexts\cite{Nagy11,Polonyi05,Pawlowski07,Blaizot11b} but few works have focused on interacting fermions.\cite{Wetterich07,Dupuis05,Kemler13} 

To set up a NPRG approach in the 2PI formalism, we add to the action $S[\psi^*,\psi]$ a regulator term $\Delta S_k$ which suppresses both fermionic and (bosonic) collective low-energy fluctuations. This allows us to define a scale-dependent 2PI effective action $\Gamma_k[\calG]$ and a scale-dependent Luttinger-Ward functional $\Phi_k[\calG]$, where $k$ is a momentum scale varying between a microscopic scale $\Lamb$ and 0. $\Delta S_k$  vanishes for $k=0$ and is chosen such that the action $S+\Delta S_\Lamb$ is noninteracting and therefore trivially solvable. With a suitable definition of $\Gamma_k[\calG]$ (slightly differing from the Legendre transform of the free energy), $\Gamma_\Lamb[\calG]$ can be made to coincide with the Hartree-Fock--RPA theory. The Luttinger-Ward functional $\Phi_{k=0}[\calG]$ of the original model is obtained from $\Phi_\Lamb[\calG]$ using a RG equation. By (approximately) solving the latter, we can then obtain the physical properties of the system we are interested in. In this 
paper we discuss 
the main properties of the 2PI-NPRG approach: i) the ``classical'' variable is the one-particle propagator $\calG$, i.e. the mean value of a composite field, and is bosonic in nature. It is itself an order parameter and there are therefore no difficulties to describe phases with spontaneously broken symmetries. It is even possible to start the RG flow in a broken-symmetry phase (with the Hartree-Fock--RPA theory as the initial condition). Moreover, the Mermin-Wagner theorem is satisfied by the renormalized theory at $k=0$. ii) The 2PI-NPRG approach is not restricted to the weak-coupling limit as is already apparent when using the Hartree-Fock--RPA theory as the initial condition of the RG flow. iii) Simple expansions of the Luttinger-Ward functional about the (possibly degenerate) equilibrium state, including only the two-particle 2PI vertex, lead to nontrivial RG equations where interactions between 
fermions and collective excitations naturally emerge. Furthermore, it is possible to parameterize the two-particle 2PI vertex by means of a small number of coupling constants. The (full, i.e. 1PI) momentum-frequency dependent two-particle vertex is computed from the 2PI vertex by solving a Bethe-Salpeter equation. Thus, contrary to the fermionic 1PI RG approach, there is no need to discretize the momentum space into patches to keep track of the momentum dependence of the two-particle 1PI vertex when solving numerically the flow equations.  

The outline of the paper is as follows. In Sec.~\ref{sec_phik} we introduce the scale dependent 2PI effective action $\Gamma_k[\calG]$ as well as the Luttinger-Ward functional $\Phi_k[\calG]$. We show how correlation functions are related to the 2PI vertices (i.e. the functional derivatives of $\Phi_k[\calG]$) and discuss the initial condition of the flow at $k=\Lamb$. In Sec.~\ref{sec_eqrg} we derive the RG equations satisfied by $\Gamma_k[\calG]$ and $\Phi_k[\calG]$, and the one- and two-particle 2PI vertices. In Sec.~\ref{sec_truncation}, we propose an approximation scheme to solve the RG equations based on a truncation of the Luttinger-Ward functional $\Phi_k[\calG]$ where only the two-particle 2PI vertex is taken into account. A summary of our approach is given in~Sec.~\ref{sec_conclu}. For clarity, many technical details are included in the Appendices.

\section{Scale-dependent Luttinger-Ward functional}
\label{sec_phik} 

\subsection{Scale-dependent 2PI effective action}

We consider a spin-$\half$ fermion system with the Euclidean action $S=S^{(0)}+\Sint$, 
\begin{equation}
\begin{split}
S^{(0)}[\psi] &= - \half \sum_{\alpha,\alpha'} \psi_\alpha \calG_{\alpha\alpha'}^{(0)-1} \psi_{\alpha'} , \\ 
\Sint[\psi] &= \frac{1}{4!} \sum_{\alpha_1\cdots\alpha_4} U_{\alpha_1\alpha_2\alpha_3\alpha_4} \psi_{\alpha_1} \psi_{\alpha_2} \psi_{\alpha_3} \psi_{\alpha_4} ,
\end{split}
\label{action0}
\end{equation} 
defined on a $d$-dimensional lattice (with $N$ sites). 
The $\psi_\alpha$'s are anticommuting Grassmann variables and the collective index $\alpha=\lbrace \r,\tau,\sig,c \rbrace$ stands for the lattice site coordinate, imaginary time, spin projection along a given axis and charge index. The latter, $c=\pm$, is such that 
\begin{equation}
\psi_\alpha = \llbrace 
\begin{array}{ccc} 
\psi_\sig(\r,\tau) & \mbox{if} & c=- , \\
\psi^*_\sig(\r,\tau) & \mbox{if} & c=+ .
\end{array}
\right.
\end{equation}
We use the notation 
\begin{equation}
\sum_\alpha = \inttau \sum_{\r,\sig,c} 
\end{equation}
with $\beta=1/T$ the inverse temperature (we set $k_B=\hbar=1$ throughout the paper). $\calG^{(0)}$ denotes the (bare) propagator and $U$  the fully antisymmetrized interaction:\cite{note1} 
\begin{equation}
\begin{gathered}
\calG^{(0)}_{\alpha\alpha'} = - \calG^{(0)}_{\alpha'\alpha} , \\ 
U_{\alpha_1\alpha_2\alpha_3\alpha_4} = - U_{\alpha_2\alpha_1\alpha_3\alpha_4} = - U_{\alpha_3\alpha_2\alpha_1\alpha_4}, \quad \mbox{etc.} 
\end{gathered}
\label{fullysym} 
\end{equation}
We assume that the action is invariant under translation, (global) SU(2) spin rotation and (global) U(1) transformation. 

To implement the RG approach, we add to the action~(\ref{action0}) the ``regulator'' term 
\begin{align}
\Delta S_k[\psi] ={}& - \half \sum_{\alpha,\alpha'} \psi_\alpha R^{(F)}_{k,\alpha\alpha'} \psi_{\alpha'} 
\nonumber \\ & + \frac{1}{4!} \sum_{\alpha_1\cdots\alpha_4} R_{k,\alpha_1\alpha_2\alpha_3\alpha_4} \psi_{\alpha_1} \psi_{\alpha_2} \psi_{\alpha_3} \psi_{\alpha_4} ,
\label{DSk} 
\end{align}
with both quadratic and quartic terms. $R^{(F)}_k$ and $R_k$ will be referred to as cutoff functions. They should be such that $\Delta S_k$ satisfies the global symmetries of the action. 

The quadratic ``fermionic'' regulator $R^{(F)}_k$ can be included in the definition of the bare propagator, 
\begin{equation}
\calG_{k,\alpha\alpha'}^{(0)-1} = \calG_{\alpha\alpha'}^{(0)-1} + R^{(F)}_{k,\alpha\alpha'} .
\end{equation}
The role of this regulator is to remove low-energy fermionic states, i.e. states near the Fermi surface. A standard choice is 
\begin{equation}
\calG^{(0)}_{k;-\sig,+\sig}(\p,i\wn) = \frac{\Theta(|\xi_\p|-\eps_k)}{i\wn-\xi_\p} 
\label{G0k}
\end{equation}
($\Theta$ denotes the step function), 
where $\wn$ is a fermionic Matsubara frequency, $\p$ the band momentum and $\eps_\p=\xi_\p+\mu$ the bare dispersion ($\mu$ is the chemical potential). $\eps_k$ is a characteristic energy scale (e.g. $\eps_k=tk^2$ in the Hubbard model with $t$ the hopping amplitude between nearest-neighbor sites) which vanishes for $k=0$ so that $\calG_{k=0}^{(0)}=\calG^{(0)}$. Instead of the sharp cutoff~(\ref{G0k}) one could choose a soft cutoff. In some cases, it is possible not to include a fermionic cutoff; for instance, when there is a spontaneously broken symmetry, the gap in the fermionic spectrum may provide a natural regulator for the fermionic degrees of freedom. In the following we shall use $\calG_{k}^{(0)}$ and not refer to $R^{(F)}_k$ anymore. We denote by $S_k^{(0)}$ the quadratic action with propagator $\calG_{k}^{(0)}$, and by $S_k$ the action $S+\Delta S_k$.

As for the ``bosonic'' quartic regulator, which modifies the fermion-fermion interaction, we require the following three properties: i) for $k$ equal  to a microscopic scale $\Lambda$ (of the order of the inverse lattice spacing), the model with action $S_\Lamb=S+\Delta S_\Lamb$ must be exactly solvable. In practice, we choose $R_\Lambda=-U$ (the system is then noninteracting),\cite{Wetterich07} which ensures that the model is exactly solvable regardless of the choice of $\calG^{(0)}_k$;\cite{note6} ii) for small $k$, $R_k$ must act as an infrared regulator for the (potentially dangerous) low-energy collective fluctuations, thus preventing any divergence in two-particle vertices and correlation functions for $k>0$; iii) $R_{k=0}$ must vanish so that the action $S_{k=0}=S$ reduces to the action~(\ref{action0}) of the original model. To relate the (exactly solvable) model with action $S_\Lambda=S_\Lamb^{(0)}$ to the model with action $S_{k=0}$, we will use a RG equation (Sec.~\ref{sec_eqrg})
. 

Although quartic regulators, modifying the fermion-fermion interaction, have been used in other works,\cite{Honerkamp04,Polonyi05,Nagy11,Streib13,Kemler13} we want to stress that $R_k$ should primarily be seen as an infrared regulator for collective fluctuations. In this respect, its $k$ dependence is expected to be crucial.  

The scale-dependent partition function reads 
\begin{equation} 
Z_k[J] = \int\calD[\psi] \,e^{ -S[\psi]-\Delta S_k[\psi] + \half \sum_{\alpha,\alpha'} \psi_\alpha J_{\alpha\alpha'} \psi_{\alpha'} } 
\label{Zdef} 
\end{equation}
in the presence of an external (antisymmetric) bilinear source $J_{\alpha\alpha'}=-J_{\alpha'\alpha}$. The one-particle propagator $\calG_{k,\gamma}=-\mean{\psi_{\alpha}\psi_{\alpha'}}$ is then obtained from 
\begin{equation}
\calG_{k,\gamma}[J] = - \frac{\delta W_k[J]}{\delta J_\gamma} ,
\label{Gdef} 
\end{equation}
where $W_k[J]=\ln Z_k[J]$ and 
\begin{equation}
\gamma = \lbrace \alpha,\alpha' \rbrace
\end{equation}
is the bosonic index obtained from the two fermionic indices $\alpha$ and $\alpha'$. Higher-order propagators are defined by 
\begin{equation}
W^{(n)}_{k,\gam_1\cdots\gam_n}[J] = \frac{\delta^n W_k[J]}{\delta J_{\gam_1}\cdots\delta J_{\gam_n}} 
\label{Wndef} 
\end{equation} 
and satisfy the symmetry properties 
\begin{equation}
\begin{split}
W^{(n)}_{k,\gam_1\cdots \gam_i \cdots \gam_j \cdots \gam_n}[J] &= W^{(n)}_{k,\gam_1\cdots \gam_j \cdots \gam_i \cdots \gam_n}[J] , \\ 
W^{(n)}_{k,\gam_1\cdots \lbrace\alpha_i,\alpha'_i\rbrace \cdots \gam_n}[J] &= - W^{(n)}_{k,\gam_1\cdots \lbrace\alpha'_i,\alpha_i\rbrace \cdots \gam_n}[J] . 
\label{Wnsym} 
\end{split}
\end{equation}

By inverting the ``equation of motion''~({\ref{Gdef}), we can express the source $J\equiv J_k[\calG]$ as a $k$-dependent functional of the propagator. We define the scale-dependent 2PI effective action
\begin{equation}
\Gamma_k[\calG] = - W_k[J] - \half \sum_\gamma J_\gamma \calG_\gamma - \frac{1}{8} \sum_{\gam_1,\gam_2} R_{k,\gam_1\gam_2} \calG_{\gam_1} \calG_{\gam_2} 
\label{gam0}
\end{equation}
(with $J\equiv J_k[\calG]$) as the Legendre transform of $W_k[J]$ to which we subtract $\frac{1}{8} \sum_{\gam_1,\gam_2} R_{k,\gam_1\gam_2} \calG_{\gam_1} \calG_{\gam_2}$. The reason for this subtraction is explained below.  

For $k=\Lambda$ the system is noninteracting, $S_\Lambda=S^{(0)}_\Lamb$, and the effective action can be computed exactly using 
\begin{align}
Z_\Lambda[J] &= \det(-\calG^{(0)-1}_\Lamb-J)^{1/2} \nonumber \\ 
&= \exp \half \tr\ln(-\calG^{(0)-1}_\Lamb-J) , 
\end{align}
where $\tr$ denotes the fermionic trace.\cite{note2} Equation~(\ref{Gdef}) then gives $\calG =(\calG^{(0)-1}_\Lamb+J)^{-1}$ and in turn 
\begin{align}
\Gamma_\Lambda[\calG] ={}& \half \tr \ln (-\calG) - \half \tr (\calG^{(0)-1}_\Lamb\calG-1) \nonumber \\ & - \frac{1}{8} \sum_{\gam_1,\gam_2} R_{\Lamb,\gam_1\gam_2} \calG_{\gam_1} \calG_{\gam_2} .
\label{gam2}  
\end{align}  

For $k<\Lambda$ the system is interacting and we write the effective action in the form 
\begin{equation}
\Gamma_k[\calG] = \half \tr \ln (-\calG) - \half \tr (\calG^{(0)-1}_k\calG-1) + \Phi_k[\calG] ,
\label{gam1}
\end{equation}
where the scale-dependent Luttinger-Ward functional $\Phi_k[\calG]$ is independent of the bare propagator $\calG_k^{(0)}$.\cite{note12} It can be defined perturbatively as the sum of the 2PI (or skeleton) diagrams with interaction vertices $U+R_k$ (i.e. diagrams that cannot be separated into two disconnected pieces by cutting at most two lines\cite{[{See, e.g., }]Haussmann99}) to which we subtract $\frac{1}{8} \sum_{\gam_1,\gam_2} R_{k,\gam_1\gam_2} \calG_{\gam_1} \calG_{\gam_2}$. Since $U+R_\Lamb=0$, all 2PI diagrams vanish and the Luttinger-Ward functional for $k=\Lamb$ reduces to
\begin{equation}
\Phi_\Lamb[\calG] = \frac{1}{8} \sum_{\gam_1,\gam_2} U_{\gam_1\gam_2} \calG_{\gam_1} \calG_{\gam_2} 
\label{Phi_init} 
\end{equation}
in agreement with~(\ref{gam2}), which reproduces the Hartree-Fock--RPA theory. We further discuss the initial condition in Sec.~\ref{subsec_init}. On the other hand, $\Phi_{k=0}[\calG]$ coincides with the usual Luttinger-Ward functional (sum of the 2PI diagrams with interaction vertex $U$) since $R_{k=0}=0$. 

The 2PI vertices are defined by the functional derivatives 
\begin{equation}
\Phi^{(n)}_{k,\gam_1\cdots\gam_n}[\calG] = \frac{\delta^n \Phi_k[\calG]}{\delta\calG_{\gam_1} \cdots \delta\calG_{\gam_n}} 
\end{equation}
and satisfy the same symmetry properties as the propagators $W^{(n)}_{k,\gam_1\cdots\gam_n}$ [Eq.~(\ref{Wnsym})]. To $m$th order in perturbation theory, they are given by 2PI diagrams with $n$ external (bosonic) legs $\gam_i$, $m$ interaction vertices $U+R_k$, and $2m-n$ propagators (these diagrams cannot be separated into two disconnected pieces by cutting at most two lines, considering every external leg $\gam_i=\lbrace\alpha_i,\alpha'_i\rbrace$ as a connected piece). Because of the definition of $\Gamma_k[\calG]$ as a slightly modified Legendre transform of $W_k[J]$ [Eq.~(\ref{gam0})], $\Phi^{(1)}_{k,\gam}$ and $\Phi^{(2)}_{k,\gam\gam'}$ include the additional contributions $-\half \sum_{\gam'} R_{k,\gam\gam'} \calG_{\gam'}$ and $-R_{k,\gam\gam'}$, respectively. 
 
\subsection{One- and two-particle propagators} 

From Eq.~(\ref{Gdef}) we deduce that $\Gamma_k[\calG]$ satisfies the ``equation of motion'' (see Appendix~\ref{app_bform})
\begin{equation}
\frac{\delta \Gamma_k[\calG]}{\delta \calG_\gam} = - J_\gam - \half \sum_{\gam'} R_{k,\gam\gam'} \calG_{\gam'}.
\label{eos} 
\end{equation} 
Together with~(\ref{gam1}), this implies Dyson's equation 
\begin{equation}
\calG^{-1}_\gam = \calG^{(0)-1}_{k,\gam} -\Sigma_{k,\gam}[\calG]
\end{equation}
with the self-energy functional 
\begin{equation}
\Sigma_{k,\gam}[\calG] = - \Phi^{(1)}_{k,\gam}[\calG] - J_\gam - \half \sum_{\gam'} R_{k,\gam\gam'} \calG_{\gam'} .
\label{self1} 
\end{equation}
Note that at this stage, $J\equiv J_k[\calG]$ is still a functional of $\calG$ whose value must be specified to obtain the propagator in the equilibrium state (Sec.~\ref{subsec_eqstate}).  

By differentiating Eq.~(\ref{eos}) wrt the source $J$ and using Eq.~(\ref{Gdef}), we obtain 
\begin{equation}
\half \sum_{\gam_3} \left( \Gamma^{(2)}_{k,\gam_1\gam_3}[\calG] + R_{k,\gam_1\gam_3} \right) W^{(2)}_{k,\gam_3\gam_2}[J]  = \calI_{\gam_1\gam_2} , 
\label{W2}
\end{equation}
where $\calI$ is the (bosonic) unit matrix defined by Eq.~(\ref{Idef}) and $\Gamma_k^{(2)}[\calG]$ the second-order functional derivative of the 2PI effective action $\Gamma_k[\calG]$. Equation~(\ref{W2}) can be rewritten as a bosonic matrix equation (see Appendix~\ref{app_bform}), 
\begin{equation}
\bigl( \Gamma_k^{(2)} + R_k \bigr)^{-1} = W_k^{(2)} . 
\label{W2inv} 
\end{equation}
From now on, in order to alleviate the notations, we suppress the $\calG$ (or $J$) dependence of the functionals $\Gamma_k^{(n)}$, $\Phi_k^{(n)}$ and $W_k^{(n)}$. From~(\ref{gam1}), we find 
\begin{equation}
\Gamma_{k,\gam_1\gam_2}^{(2)} = \Pi^{-1}_{\gam_1\gam_2} +\Phi^{(2)}_{k,\gam_1\gam_2} ,
\label{Gam2}
\end{equation}
where 
\begin{equation}
\Pi^{-1}_{\gam_1\gam_2} = - \calG^{-1}_{\alpha_1\alpha_2} \calG^{-1}_{\alpha'_1\alpha'_2} + \calG^{-1}_{\alpha_1\alpha'_2} \calG^{-1}_{\alpha'_1\alpha_2}
\end{equation} 
is the inverse (in a bosonic matrix sense) of the pair propagator $\Pi\equiv\Pi[\calG]$, 
\begin{equation}
\Pi_{\gam_1\gam_2} = - \calG_{\alpha_1\alpha_2} \calG_{\alpha'_1\alpha'_2} + \calG_{\alpha_1\alpha'_2} \calG_{\alpha'_1\alpha_2} .
\end{equation} 
This allows us to rewrite the bosonic propagator $W^{(2)}_k$ as 
\begin{equation}
W^{(2)}_k = (\Pi^{-1} + \calX_k )^{-1} = \Pi - \Pi \calX_k W^{(2)}_k ,
\label{W2a}
\end{equation}
where 
\begin{equation}
\calX_k = \Phi^{(2)}_k + R_k . 
\end{equation}
$W_k^{(2)}= \Pi-\Pi\calY_k\Pi$ can also be related to the 1PI two-particle vertex $\calY_k$
\begin{equation}
\calY_k = (\calX_k^{-1}+\Pi)^{-1} = \calX_k - \calX_k \Pi \calY_k . 
\label{Ydef}
\end{equation}
Equations~(\ref{W2a}) and (\ref{Ydef}) are Bethe-Salpeter equations relating $W_k^{(2)}$ and $\calY_k$ to the (regularized) 2PI vertex $\calX_k=\Phi_k^{(2)}+R_k$. Equation~(\ref{W2a}) can also be seen as a Dyson equation for the (bosonic) pair propagator $W_k^{(2)}$, with bare propagator $\Pi$ and ``self-energy'' $\calX_k=\Phi_k^{(2)}+R_k$. Since $W_k^{(2)-1}=\Pi^{-1}+\Phi^{(2)}_k+R_k$, we see that $R_k$ naturally appears as a regulator for the bosonic fluctuations.

\subsection{Particle-particle and particle-hole channels}
\label{subsec_ppph} 

So far we have used a compact notation where all fermionic indices are gathered in the collective index $\alpha$. {\it In fine} it is however necessary to explicitly introduce the singlet and triplet particle-particle (pp) channels, as well as the charge and spin particle-hole (ph) channels. To this end we define the matrices 
\begin{equation}
\tau^\nu_{--} = \tau^{\nu\dagger}_{++} = i \sig^\nu \sig^y , \qquad
\tau^\nu_{-+} = \tau^{\nu T}_{+-} =  \sig^\nu , 
\end{equation}
($\nu=0,x,y,z$), where $\sig^0$ is the $2\times 2$ unit matrix and $\sig^x$, $\sig^y$, $\sig^z$ the Pauli matrices. They satisfy the property
\begin{equation}
\half \tr \bigl(\tau_{cc'}^{\nu\dagger} \tau_{cc'}^{\nu'} \bigr) = \delta_{\nu,\nu'}. 
\end{equation}
In the following we use the notation $x=(\r,\tau)$ and $X=(x,\sig)$ so that $\psi_+(X)=\psi^*_\sig(x)$ and $\psi_-(X)=\psi_\sig(x)$. To alleviate the notations, we drop the index $k$ in this section. 

Let first consider the composite field $O_{cc'}(X,X') = \psi_c(X)\psi_{c'}(X')$. It can be decomposed on various spin channels using 
\begin{equation}
O_{cc'}(X,X') = \half \sum_\nu \bigl(\tau^\nu_{cc'}\bigr)_{\sig\sigp} O_{cc'}^\nu(x,x') ,
\end{equation}
where
\begin{equation}
O_{cc'}^\nu(x,x') = \sum_{\sig,\sig'} \bigl(\tau^{\nu\dagger}_{cc'}\bigr)_{\sigp\sig} O_{cc'}(X,X') .
\end{equation}
In the ph channel ($c=-c'$), $\nu=0$ corresponds to the charge component and $\nu=x,y,z$ to the three spin components, while in the pp channel $\nu=0$ and $\nu=x,y,z$ correspond to the singlet and triplet components, respectively.\cite{note7} 

The Fourier transformed field is defined by 
\begin{equation}
O_{cc'}^\nu(x,x') = \frac{1}{\beta N} \sum_{p,p'} e^{-i(cpx+c'p'x')} O_{cc'}^\nu(p,p') , 
\end{equation}
where $p=(\p,i\wn)$ with $\wn$ a fermionic Matsubara frequency and $px=\p\cdot\r-\wn\tau$. It is convenient to introduce the total and relative momentum-frequency of the pair,
\begin{equation}
q = p'+ cc'p, \qquad l = \half(p'-cc'p) , 
\label{qldef1}
\end{equation}
which allows us to define the following pp and ph composite fields, 
\begin{equation}
\begin{split}
O^\nu_{\rm pp}(q,l) &= O^\nu_{--}(q,l) , \\ 
O^{\nu\dagger}_{\rm pp}(q,l) &= O^\nu_{++}(q,-l) , \\ 
O^\nu_{\rm ph}(q,l) &= O^\nu_{+-}(q,l) .
\end{split}
\label{qldef}
\end{equation}
These operators are particularly useful when dealing with the pair propagator $W_k^{(2)}$ or the 1PI vertex $\calY_k$.

\subsubsection{Propagators}

The one-particle propagator reads 
\begin{align}
\calG_{cc'}(X,X') &= - \mean{O_{cc'}(X,X')} \nonumber \\   
&= \half \sum_\nu \bigl(\tau^\nu_{cc'}\bigr)_{\sig\sigp} \calG_{cc'}^\nu(x,x') , 
\label{Gcc}
\end{align}
where 
\begin{equation}
\calG_{cc'}^\nu(x,x') = - \mean{O_{cc'}^\nu(x,x')} .
\end{equation} 
Using more standard notations, we introduce the normal and anomalous propagators 
\begin{equation}
\begin{split}
G_{\sig\sig'}(x,x') &= - \mean{\psi_\sig(x) \psi^*_{\sig'}(x')} = \calG_{-+}(X,X') , \\ 
F_{\sig\sig'}(x,x') &= - \mean{\psi_\sig(x) \psi_{\sig'}(x')} = \calG_{--}(X,X') , \\ 
F^\dagger_{\sig\sig'}(x,x') &= - \mean{\psi^*_\sig(x) \psi^*_{\sig'}(x')} = \calG_{++}(X,X'). 
\end{split}
\end{equation}
The normal propagator can be decomposed into charge and spin components, 
\begin{equation}
G_{\sig\sig'}(x,x') = \half \sig^0_{\sig\sig'} \Gch(x,x') + \half \sigmabf_{\sig\sig'} \cdot \Gspv(x,x') ,
\end{equation}
where $\Gch=\calG_{-+}^0$ and $\Gsp^\nu=\calG_{-+}^\nu$ ($\nu=x,y,z$). The anomalous propagator can be decomposed into singlet and triplet components, 
\begin{equation}
\begin{split}
F_{\sig\sig'}(x,x') &= \half (i\sig_y)_{\sig\sig'} \Fs(x,x') +  \half (i\sigmabf \sig_y)_{\sig\sig'} \cdot \Ftv(x,x'), \\ 
F^\dagger_{\sig\sig'}(x,x') &= \half (i\sig_y)^\dagger_{\sig\sig'} \Fs^\dagger(x,x') +  \half (i\sigmabf \sig_y)^\dagger_{\sig\sig'} \cdot \Ftv^\dagger(x,x') ,
\end{split}
\end{equation}
where $\Fs=\calG_{--}^0$, $\Fs^\dagger=\calG_{++}^0$, $\Ft^\nu=\calG_{--}^\nu$ and $\Ft^{\nu\dagger}=\calG_{++}^\nu$ ($\nu=x,y,z$). 

The two-particle propagator is defined by 
\begin{multline}
W^{(2)}_{c_1c_1'c_2c_2'}(X_1,X_1';X_2,X_2') \\ =
\mean{O_{c_1c_1'}(X_1,X_1') O_{c_2c_2'}(X_2,X_2')}_c 
\end{multline}
or
\begin{align}
W^{(2)\nu_1\nu_2}_{c_1c_1'c_2c_2'}(x_1,x'_1;x_2,x'_2) &= \mean{O^{\nu_1}_{c_1c_1'}(x_1,x_1') O^{\nu_2}_{c_2c_2'}(x_2,x_2')}_c ,
\end{align}
where $\mean{OO'}_c=\mean{OO'}-\mean{O}\mean{O'}$ 
(similar expressions hold for the pair propagator $\Pi$). When the global U(1) invariance is not spontaneously broken, we can distinguish between pp ($c_1=c_1'=-c_2=-c'_2$) and ph ($c_1+c_1'=c_2+c_2'=0$) propagators, 
\begin{equation}
\begin{split}
W^{(2)\nu_1\nu_2}_{\rm pp}(x_1,x'_1;x_2,x'_2) &= W^{(2)\nu_1\nu_2}_{--++}(x_1,x'_1;x_2,x'_2) , \\ 
W^{(2)\nu_1\nu_2}_{\rm ph}(x_1,x'_1;x_2,x'_2) &= W^{(2)\nu_1\nu_2}_{+-+-}(x_1,x'_1;x_2,x'_2) 
\end{split}
\end{equation}
(and similarly for $\Pi$). When the global U(1) symmetry is broken, there are other nonzero propagators such as $W^{(2)}_{----}$ and $W^{(2)}_{+---}$. Finally, introducing the total and relative momentum-frequency of the pair [Eq.~(\ref{qldef1})], we define the propagators $W^{(2)\nu_1\nu_2}_{c_1c_1'c_2c_2'}(q_1,l_1;q_2,l_2)$ and $\Pi^{\nu_1\nu_2}_{c_1c_1'c_2c_2'}(q_1,l_1;q_2,l_2)$ (see Appendix~\ref{app_bseq} for a more detailed discussion).  

\subsubsection{Vertices}  

A similar decomposition holds for the vertices, 
\begin{equation}
\begin{split}
\Phi^{(1)}_{cc'}(X,X') = \sum_\nu \bigl(\tau^{\nu\dagger}_{cc'}\bigr)_{\sigp\sig} \Phi^{(1)\nu}_{cc'}(x,x') , \\  \Phi^{(1)\nu}_{cc'}(x,x') = \half \sum_{\sig,\sigp} \bigl(\tau^\nu_{cc'}\bigr)_{\sig\sigp} 
\Phi^{(1)}_{cc'}(X,X')  
\end{split}
\end{equation}
and analog expressions for the two-particle vertices $\Phi^{(2)}_{c_1c_1'c_2c_2'}$, $\calX_{c_1c_1'c_2c_2'}$ and $\calY_{c_1c_1'c_2c_2'}$ (see Appendix~\ref{app_bseq}). We define 
\begin{equation} 
\begin{split}
\Phi^{(1)\nu}_{\rm pp}(x,x') &= \Phi^{(1)\nu}_{++}(x,x') , \\ 
\Phi^{(1)\nu}_{\rm ph}(x,x') &= \Phi^{(1)\nu}_{+-}(x,x') 
\end{split}
\label{phi1a}
\end{equation}
in the pp and ph channels, respectively. When the global U(1) symmetry is not broken, it is convenient to introduce 
\begin{equation} 
\begin{split}
\Phi^{(2)\nu_1\nu_2}_{\rm pp}(x_1,x'_1;x_2,x'_2) &= \Phi^{(2)\nu_1\nu_2}_{++--}(x_1,x'_1;x_2,x'_2) , \\ 
\Phi^{(2)\nu_1\nu_2}_{\rm ph}(x_1,x'_1;x_2,x'_2) &= \Phi^{(2)\nu_1\nu_2}_{+-+-}(x_1,x'_1;x_2,x'_2) 
\end{split}
\end{equation}
(and similarly for $\calX$ and $\calY$). When the global U(1) symmetry is broken, there are additional (nonzero) vertices such as $\Phi^{(2)\nu_1\nu_2}_{----}$ and $\Phi^{(2)\nu_1\nu_2}_{---+}$.

\subsubsection{Bethe-Salpeter equations} 
\label{subsubsec_bseq} 

We are now in a position to write the Bethe-Salpeter equations~(\ref{W2a}) and (\ref{Ydef}) in a more explicit form (Appendix~\ref{app_bseq}):
\begin{multline}  
\calY^{\nu_1\nu_2}_{c_1c'_1c_2c'_2}(q_1,l_1;q_2,l_2) = \calX^{\nu_1\nu_2}_{c_1c'_1c_2c'_2}(q_1,l_1;q_2,l_2)
 \\  - \frac{1}{4\beta N} \sum_{c_3\cdots c'_4 \atop \nu_3,\nu_4} \sum_{q_3,l_3,q_4,l_4}  \calX^{\nu_1\nu_3}_{c_1c'_1c_3c'_3}(q_1,l_1;q_3,l_3) \\ \times \Pi^{\nu_3\nu_4}_{c_3c'_3c_4c'_4}(q_3,l_3;q_4,l_4) \calY^{\nu_4\nu_2}_{c_4c'_4c_2c'_2}(q_4,l_4;q_2,l_2) 
\label{eqbs1}
\end{multline}
and 
\begin{multline}
W^{(2)\nu_1\nu_2}_{c_1c'_1c_2c'_2}(q_1,l_1;q_2,l_2) = \Pi^{\nu_1\nu_2}_{c_1c'_1c_2c'_2}(q_1,l_1;q_2,l_2)
 \\  - \frac{1}{4\beta N} \sum_{c_3\cdots c'_4 \atop \nu_3,\nu_4} \sum_{q_3,l_3,q_4,l_4}  \Pi^{\nu_1\nu_3}_{c_1c'_1c_3c'_3}(q_1,l_1;q_3,l_3) \\ \times \calX^{\nu_3\nu_4}_{c_3c'_3c_4c'_4}(q_3,l_3;q_4,l_4) W^{(2)\nu_4\nu_2}_{c_4c'_4c_2c'_2}(q_4,l_4;q_2,l_2) .
\label{eqbs1a}
\end{multline}
When the global U(1) symmetry is not broken, we can consider separately the pp and ph channels.

\subsection{Equilibrium state} 
\label{subsec_eqstate} 

Why do we consider the functional $\Gamma_k[\calG]$ [Eq.~(\ref{gam0})] rather than the true Legendre transform of $W_k[J]$? The reason is that $S_k=S+\Delta S_k$ can be significantly different from the action $S$ of the model we are interested in. This is most notably true when $k=\Lamb$ since the system is noninteracting ($U+R_\Lamb=0$). When considering $\Gamma_k[\calG]$, we partially compensate the effect of $R_k$ by subtracting $\frac{1}{8} \sum_{\gam_1,\gam_2} R_{k,\gam_1\gam_2} \calG_{\gam_1} \calG_{\gam_2}$. The compensation is exact at the Hartree-Fock level where the Luttinger-Ward functional is truncated to order $\calO(U+R_k)$. Thus the 
main difference between $\Gamma_{k=0}[\calG]$ and $\Gamma_{k}[\calG]$ is that the latter takes into account low-energy (wrt the momentum scale $k$) bosonic fluctuations only at the Hartree-Fock level. We can therefore interpret $\Gamma_k[\calG]$ as a coarse-grained free energy with a coarse-graining length scale of order $k^{-1}$.\cite{note3} 

Consequently, we define the propagator $\bar\calG_k$ in the equilibrium state from the minimum of $\Gamma_k[\calG]$,  
\begin{equation}
\frac{\delta \Gamma_k[\calG]}{\delta \calG_\gam} \biggl|_{\bar\calG_k} = 0 .
\label{Gbardef} 
\end{equation}
Note that this amounts to evaluating the equation of motion~(\ref{self1}) with an external source $J_{k,\gam}[\bar\calG]=-\half \sum_{\gam'}R_{k,\gam\gam'} \bar\calG_{\gam'}$. The corresponding self-energy coincides with the one-particle 2PI vertex, 
\begin{equation}
\bar\Sigma_{k,\gam} = - \bar\Phi^{(1)}_{k,\gam} ,
\label{self2}
\end{equation}
where we use the notation $\bar\Phi^{(n)}_{k,\gam} \equiv \Phi^{(n)}_{k,\gam}[\bar\calG_k]$. If some global symmetries are spontaneously broken then the equilibrium state is degenerate.
As in the case of the one-particle 2PI vertex [Eqs.~(\ref{phi1a})], we can define ph and pp components of the self-energy, 
\begin{equation}
\begin{split}
\bar\Sig_{k,\rm ph}^\nu(x,x') &= \bar\Sig_{k,+-}^\nu(x,x') , \\ 
\bar\Sig_{k,\rm pp}^\nu(x,x') &= \bar\Sig_{k,++}^\nu(x,x') ,
\end{split}
\end{equation}
where $\bar\Sig_{k,\rm ph}^0=\bar\Sig_{k,\rm ch}$, $\bar\Sig_{k,\rm ph}^\nu=\bar\Sig^\nu_{k,\rm sp}$, $\bar\Sig_{k,\rm pp}^0=\bar\Sig_{k,\rm s}$ and $\bar\Sig_{k,\rm pp}^\nu=\bar\Sig^\nu_{k,\rm t}$ ($\nu=x,y,z$).   

A similar reasoning holds for the two-particle correlation function. Collective modes and other two-particle properties in the equilibrium state should be obtained from the propagator $\bar\Gamma^{(2)-1}_k$ rather than $\bar W_k^{(2)}=(\bar\Gamma^{(2)}_k+R_k)^{-1}$. For $k=0$, both correlation functions coincide: $\bar W_{k=0}^{(2)}=\bar\Gamma^{(2)-1}_{k=0}$.

\subsection{Initial condition: Hartree-Fock--RPA theory}
\label{subsec_init}  

From~(\ref{Phi_init}), we deduce that for $k=\Lamb$ the one- and two-particle 2PI vertices are given by  
\begin{equation}
\begin{gathered}
\Phi^{(1)}_{\Lamb,\gam} = \frac{1}{2} \sum_{\gam'} U_{\gam\gam'} \calG_{\gam'} , \\ 
\Phi^{(2)}_{\Lamb,\gam\gam'} =  U_{\gam\gam'} ,
\end{gathered}
\label{phi12}
\end{equation}
with all higher-order 2PI vertices vanishing. Equation~(\ref{self2}) then yields the self-energy equation  
\begin{equation}
\bar\Sigma_{\Lamb,\gam} = - \half \sum_{\gam'} U_{\gam\gam'} \bar\calG_{\Lambda,\gam'} , 
\label{self3}
\end{equation}
which corresponds to a generalized (i.e. with possible broken symmetries) Hartree-Fock approximation. If we use the sharp fermionic cutoff~(\ref{G0k}) with $\eps_\Lamb\geq\mbox{max}_\p |\xi_\p|$, all fermionic degrees of freedom are suppressed when $k=\Lamb$ and both $\bar\Sig_{\Lamb,\gam}$ and $\bar\calG_{\Lamb,\gam}$ vanish. In the following we discuss the opposite case where no fermionic regulator is included in the action ($\calG^{(0)}_k=\calG^{(0)}$). Equation~(\ref{self3}) then coincides with the standard Hartree-Fock approximation. In the case of a broken-symmetry state, it yields the self-consistent (mean-field) equation for the order parameter (see below the discussion of the Hubbard model).  

Since $U+R_\Lamb=0$, $\bar W_k^{(2)}\equiv W^{(2)}_k[\bar\calG_k]$ is equal to $\bar\Pi_k\equiv\Pi[\bar\calG_k]$. On the other hand, the correlation function $\bar\Gamma_k^{(2)-1}$ (see the discussion at the end of Sec.~\ref{subsec_eqstate}) is obtained from a Bethe-Salpeter equation with Hartree-Fock propagators $\bar\calG_\Lamb$ and bare interaction vertex $\bar\Phi^{(2)}_\Lamb=U$, which corresponds to the RPA. This approximation is conserving in the sense of Baym and Kadanoff.\cite{Baym61,Baym62} In particular, in the case of a spontaneously broken continuous symmetry, it satisfies the Goldstone theorem (see Sec.~\ref{subsec_goldstone} for a further discussion of the Goldstone theorem). 

\subsubsection*{Two-dimensional Half-filled Hubbard model}

In this section we explicit the initial condition of the RG flow in the half-filled Hubbard model defined on a square lattice in the absence of a fermionic regulator (see Appendix~\ref{app_init} for more details). The initial value~(\ref{Phi_init}) of the Luttinger-Ward functional reads ($U$ denotes the on-site interaction)
\begin{equation}
\Phi_\Lambda[\calG] = U \inttau \sum_\r \Bigl[ |\Delta(x)|^2 + \quarter \rho(x)^2 - \quarter \S(x)^2 \Bigl] ,
\label{Phi_init_HM}  
\end{equation}
where 
\begin{equation}
\begin{gathered} 
\Delta(x) = \mean{\psi_\down(x) \psi_\up(x)} = \half F_s(x,x) , \\ 
\Delta(x)^* = \mean{\psi_\up^*(x) \psi_\down^*(x)} = \half F^\dagger_s(x,x) , \\ 
\rho(x) = \sum_\sig \mean{\psi^*_\sig(x) \psi_\sig(x)} = \Gch(x,x^+)  , \\ 
S^\nu(x) = \sum_{\sig,\sig'} \mean{\psi^*_\sig(x) \sig^\nu_{\sig\sig'} \psi_{\sig'}(x)}  = \Gsp^\nu(x,x^+), 
\end{gathered}
\end{equation}
are the (singlet) superconducting order parameter, the charge and spin densities, respectively. We use the notation $x=(\r,\tau)$ and $x^+=(\r,\tau+0^+)$. Equation~(\ref{Phi_init_HM}) is most simply obtained by using $\Phi_\Lambda[\calG]\equiv \mean{\Sint}_{\rm HF}$ where $\Sint=U\inttau \sum_\r \psi^*_\up \psi^*_\down \psi_\down \psi_\up$ and $\mean{\Sint}_{\rm HF}$ is computed using Wick's theorem (Hartree-Fock approximation).  

The 2PI vertex is given by the bare interaction [Eq.~(\ref{phi12})]:
\begin{equation}
\begin{split}
U_{\Lamb,\rm ch} &= U^{00}_{+-+-} = U/2, \\
U_{\Lamb,\rm sp} &= U^{\nu\nu}_{+-+-} = - U/2 \quad (\nu\neq 0)
\end{split}
\label{Udef1}
\end{equation}
in the ph channel, and 
\begin{equation}
\begin{split}
U_{\Lamb,\rm s} &= U^{00}_{++--} = U, \\
U_{\Lamb,\rm t} &= U^{\nu\nu}_{++--} = 0 \quad (\nu\neq 0)
\end{split}
\label{Udef2}
\end{equation}
in the pp channel. To alleviate the notations, we  now drop the $\Lamb$ index (i.e. $U_{\Lamb,\rm ch}\equiv \Uch$, etc.). 

As for the self-energy $\bar\Sig\equiv \bar\Sig_\Lamb$ [Eq.~(\ref{self3})], we obtain 
\begin{equation}
\bar\Sigma_{\rm s}(x,x') = - \delta(x-x') \frac{\Us}{2} \bFs(x,x) ,
\label{gap1} 
\end{equation}
and  
\begin{equation}
\begin{split}
\bar\Sig_{\rm ch}(x,x') &= \delta(x-x'{}^+) \Uch \bGch(x,x^+) , \\ 
\bar\Sig^\nu_{\rm sp}(x,x') &= \delta(x-x'{}^+) \Usp \bGsp^\nu(x,x^+) .
\label{gap1a} 
\end{split}
\end{equation}
At half-filling $\bGch(x,x^+)=1$ and the Hartree-Fock self-energy in the charge channel is given by $\bar\Sig_{\rm ch}(x,x') = \delta(x-x'{}^+) \Uch$. Since $\mu=U/2$ due to the ph symmetry of the half-filled model, $\bar\Sig_{\rm ch}$ cancels the chemical potential. In the following, we include the charge self-energy by setting $\mu=0$. In the attractive Hubbard model ($U<0$), $\bar\Sig_{\rm sp}^\nu$ vanishes and we recover the BCS theory with $\Delta(x)$ the order parameter and Eq.~(\ref{gap1}) as the (mean-field) gap equation. In the repulsive model, the anomalous self-energy $\bar\Sig_{\rm s}$ vanishes, and Eq.~(\ref{gap1a}) is the gap equation for the magnetic order parameter $S^\nu(x)$. 

We now focus on the repulsive case. Assuming an antiferromagnetic state polarized along the $z$ axis, 
\begin{equation}
\bar\Sig^\nu_{\rm sp}(x,x') = - \delta_{\nu,z} \delta(x-x'{}^+) m (-1)^\r ,
\label{gap1b} 
\end{equation}
which leads to the standard mean-field expressions of the propagators $\bGch$ and $\bGsp$ (Appendix~\ref{app_init}). The antiferromagnetic order parameter $m$ satisfies the gap equation 
\begin{equation}
\frac{m}{U} = T \sum_{\wn} \int \frac{d^2p}{(2\pi)^2} \frac{m}{\wn^2+\Ep^2} , 
\label{gap2}
\end{equation}
where $\Ep=(\xip^2+m^2)^{1/2}$ and $\xip=\eps_\p-\mu=\eps_\p$ ($\wn$ denotes a fermionic Matsubara frequency). The momentum integration is restricted to the first Brillouin zone of the reciprocal lattice. $m$ is nonzero below the (mean-field) transition temperature $T_c^{\rm HF}$.  In the limit $U\gg t$, Eq.~(\ref{gap2}) yields $T_c^{\rm HF}\simeq U/4$. 

The dispersion of the collective modes is obtained from the poles of $\bar\Gamma^{(2)-1}$ after analytic continuation to real frequencies. At zero temperature, this yields a (Goldstone) spin-wave mode with (square) velocity
\begin{equation} 
c^2 = \frac{\llangle\frac{\eps^2}{E^3}\rrangle  \llangle\frac{\eps_1^2}{E^3}\rrangle}{\llangle \frac{1}{E^3} \rrangle\llangle \frac{1}{E} \rrangle } , 
\label{velocity} 
\end{equation}
which is the known RPA result (Appendix~\ref{app_init}).\cite{Schrieffer89,Singh90,Chubukov92,Borejsza04} Here we use the notation $\eps=\eps_\p$, $\eps_1=\partial_{p_x}\eps$, $\eps_2=\partial^2_{p_x}\eps$, $E=\Ep$, and $\langle\cdots\rangle$ denotes a momentum integration. In the large $U$ limit, we find $c=\sqrt{2}J$ ($J=4t^2/U$), which agrees with the result obtained from the Heisenberg model in the spin-wave approximation.\cite{Auerbach_book}

A nice feature of the Hartree-Fock--RPA theory about the antiferromagnetic state is that it captures some aspects of the strong correlations in the large $U$ limit (as shown above, it provides us with a good estimate of the spin-wave mode velocity $c\sim J$). This is a direct consequence of the fermionic self-energy (i.e. the gap $m$ at the Hartree-Fock level) included in the propagator in the broken-symmetry phase. Nevertheless, the Hartree-Fock--RPA theory fails in two dimensions since it predicts long-range antiferromagnetic order at finite temperature in contradiction with the Mermin-Wagner theorem. It is also difficult to incorporate within this approach the feedback of collective fluctuations on the fermionic excitations. We shall see in the following that these shortcomings are expected to be overcome by the 2PI-NPRG approach.

The initial condition of the RG flow is different if we include a fermionic regulator $R_k^{(F)}$. As previously pointed out, $\bar\Sig_{\Lamb,\gam}$ vanishes with the regulator~(\ref{G0k}). In this case, the RG flow starts in the normal phase. At half filling, the spin part $\bar\Sig_{k,\rm sp}^\nu$ of the self-energy becomes nonzero in the course of the flow thus signaling that the ground state is antiferromagnetic. The conclusion that the NPRG captures some aspects of the strong correlations in the large $U$ limit is however unchanged; the NPRG will give a spin-wave mode velocity $c_{k=0}$ of order $J$ ($c_{k=0}$ will differ from $\sqrt{2}J$ since fluctuations beyond spin-wave theory are taken into account by the RG equation).

\subsection{Goldstone's theorem}
\label{subsec_goldstone} 

The possibility to describe phases with spontaneously broken continuous symmetries implies that we have to deal with gapless Goldstone bosons. To see how this works, let us again consider the antiferromagnetic phase of the two-dimensional Hubbard model at half-filling. We introduce the spin-spin correlation function 
\begin{align}
W^{(2)\nu_1\nu_2}_{k,\rm sp}(x_1,x_2) &= W^{(2)\nu_1\nu_2}_{k,\rm sp}(x^+_1,x_1;x^+_2,x_2) \nonumber \\ 
&\equiv W^{(2)\nu_1\nu_2}_{k,+-+-}(x^+_1,x_1;x^+_2,x_2)
\label{Wsp} 
\end{align}
where $\nu_1,\nu_2=x,y,z$, and assume the magnetic order to be polarized along the $z$ axis. In the transverse spin channel, there are two independent nonzero correlation functions,
\begin{equation}
\begin{gathered}
\bar W^{(2)xx}_{k,\rm sp}(q,q) = \bar W^{(2)yy}_{k,\rm sp}(q,q) , \\
\bar W^{(2)xy}_{k,\rm sp}(q,q+Q) , 
\end{gathered}
\end{equation}
where $Q=(\Q,0)$ with $\Q=(\pi,\pi)$. Since long-range order takes place in the $s$-wave ph channel, we can take the cutoff function
\begin{align}
R_{k,\rm sp}^{\nu_1\nu_2}(x_1,x'_1;x_2,x'_2) ={}& \delta_{\nu_1,\nu_2} \delta(x_1-x'_1{}^+) \delta(x_2-x'_2{}^+) \nonumber \\ & \times R_{k,\rm sp}(x_1-x_2) .
\end{align}
Spin-rotation invariance implies the Ward identity (see Appendix~\ref{app_goldstone})
\begin{equation}
\bar W^{(2)xx}_{k,\rm sp}(Q,Q) = \bar W^{(2)yy}_{k,\rm sp}(Q,Q) = \frac{1}{R_{k,\rm sp}(Q)} . 
\end{equation}
Only for $k=0$ do we recover the standard form of the Goldstone theorem,
\begin{equation}
\lim_{q\to Q} \bar W^{(2)xx}_{k=0,\rm sp}(q,q)=\infty ,
\end{equation}
since $R_{k=0}$ vanishes. The presence of a gap in the Goldstone modes at finite $k$ is a consequence of defining the equilibrium state from the minimum of $\Gamma_k[\calG]$ and not the true Legendre transform (which amounts to solving the Dyson equation with a nonzero external source; see Sec.~\ref{subsec_eqstate}). Again we see that $R_k$ acts as an infrared regulator for the collective (bosonic) fluctuations.

\section{Exact RG equation} 
\label{sec_eqrg} 

In this section, we derive the exact RG equations for the 2PI effective action $\Gamma_k[\calG]$ and the Luttinger-Ward functional $\Phi_k[\calG]$. We also consider the one-particle and two-particle 2PI vertices, which play a crucial role in the approximation scheme proposed in Sec.~\ref{sec_truncation}, and the thermodynamic potential.

\subsection{Luttinger-Ward functional} 

The derivation of the RG equation for the 2PI effective action follows the standard approach in the 1PI formalism.\cite{Berges02,Delamotte12,Kopietz_book} From Eq.~(\ref{Zdef}) we obtain a RG equation for the free energy, 
\begin{align} 
\dk W_k[J] ={}& \frac{1}{2 Z_k[J]} \sum_\gam \dot\calG^{(0)-1}_{k,\gam}  \frac{\delta Z_k[J]}{\delta J_{\gam}} \nonumber \\ &  - \frac{1}{4!Z_k[J]}  \sum_{\gam_1,\gam_2} \dot R_{k,\gam_1\gam_2} \frac{\delta^2 Z_k[J]}{\delta J_{\gam_1}\delta J_{\gam_2}} ,
\label{dkW}
\end{align}
where the derivative is taken at fixed external source $J$ and the dot denotes a $k$ derivative (e.g. $\dot R_k=\partial_k R_k$). Equation~(\ref{dkW}) can be rewritten as 
\begin{multline} 
\dk W_k[J] = \half \sum_\gam \dot \calG^{(0)-1}_{k,\gam}  W_{k,\gam}^{(1)}[J]  
 - \frac{1}{4!}  \sum_{\gam_1,\gam_2} \dot R_{k,\gam_1\gam_2}  \\ \times \Bigl( W^{(2)}_{k,\gam_1\gam_2}[J]
 + W^{(1)}_{k,\gam_1}[J]W^{(1)}_{k,\gam_2}[J] \Bigr) .
\end{multline}

Considering now the 2PI effective action~(\ref{gam0}), we obtain    
\begin{equation}
\dk \Gamma_k[\calG] = - \dk W_k[J]\Bigl|_J  - \frac{1}{8} \sum_{\gam_1,\gam_2} \dot R_{k,\gam_1\gam_2} \calG_{\gam_1} \calG_{\gam_2} 
\end{equation}
(we have used (\ref{Gdef})), where the derivative is taken at fixed propagator $\calG$. Using~(\ref{dkW}), we finally obtain 
\begin{align}
\dk \Gamma_k[\calG] ={}& - \half \tr \bigl( \dot \calG^{(0)-1}_k \calG \bigr) \nonumber \\ & + \frac{1}{3!} \Tr \llbrace \dot R_k \bigl(\Gamma_k^{(2)}+R_k \bigr)^{-1} \rrbrace \nonumber \\ & 
- \frac{1}{12} \sum_{\gam_1,\gam_2} \dot R_{k,\gam_1\gam_2} \calG_{\gam_1} \calG_{\gam_2} ,
\label{dkGam1} 
\end{align}
where $\tr$ and $\Tr$ denote fermionic and bosonic traces, respectively.\cite{note2,note8} The first term in the rhs of~(\ref{dkGam1}) is the one obtained in the 1PI formalism (with $\dk \calG^{(0)-1}_k=\dk R^{(F)}_k$).\cite{Wetterich93} The second one has a similar structure but at the level of collective fluctuations since it involves the bosonic propagator $W_k^{(2)}=(\Gamma_k^{(2)}+R_k)^{-1}$.

\begin{figure}
\centerline{\includegraphics[width=6.cm]{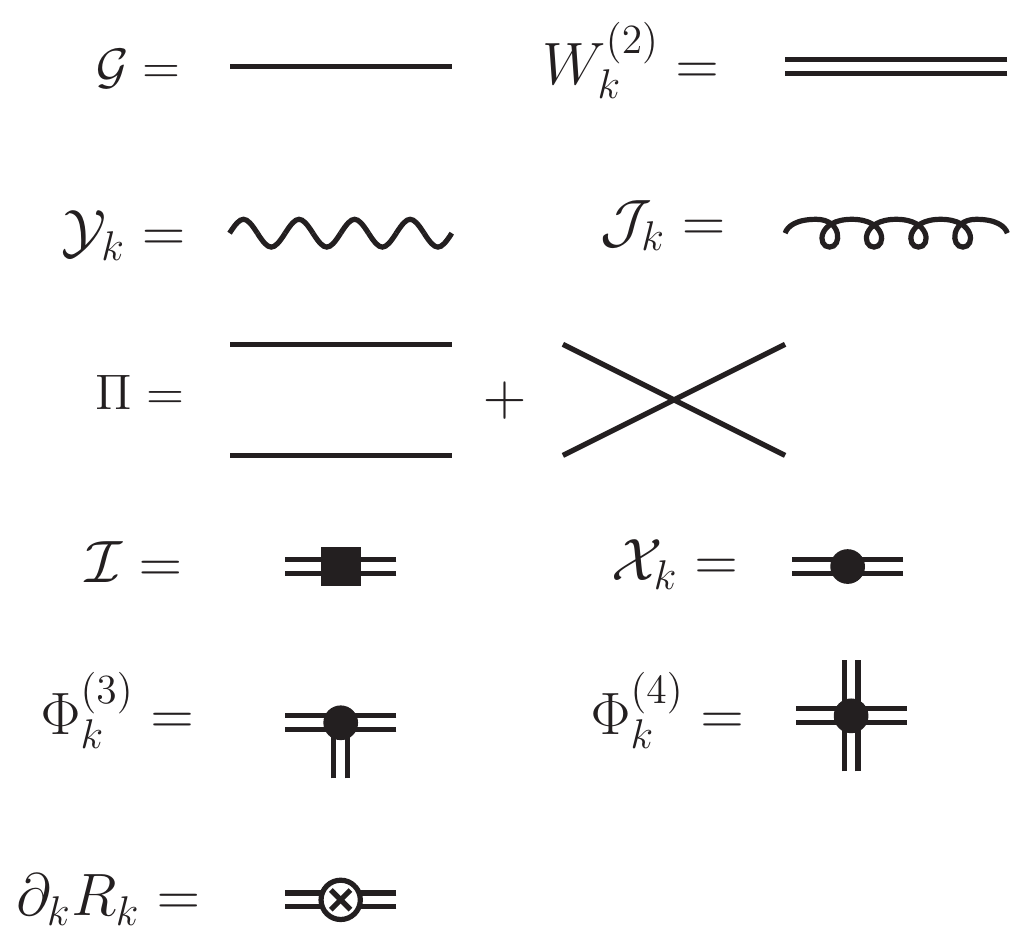}}
\caption{Diagrammatic representation of propagators and vertices.}
\label{fig_symbols}
\end{figure}

Using~(\ref{Gam2}) and the relation~(\ref{gam1}) between the 2PI effective action and the Luttinger-Ward functional, we deduce 
\begin{equation}
\dk \Gamma_k[\calG] = - \half \tr \bigl( \dot \calG^{(0)-1}_k \calG \bigr) + \dk \Phi_k[\calG]
\label{rgeq6} 
\end{equation}
and 
\begin{align}
\dk \Phi_k[\calG] ={}& \frac{1}{3!}  \Tr \llbrace \dot R_k \bigl(\Pi^{-1}+\Phi_k^{(2)}+R_k \bigr)^{-1} \rrbrace  \nonumber \\ 
& - \frac{1}{12} \sum_{\gam_1,\gam_2} \dot R_{k,\gam_1\gam_2} \calG_{\gam_1} \calG_{\gam_2} .
\label{rgeq1}
\end{align} 
Equation~(\ref{rgeq1}) is shown diagrammatically in Fig.~\ref{fig_rgeq} (the diagrammatic representation of the propagators and vertices, as well as the Bethe-Salpeter equations, are shown in Figs.~\ref{fig_symbols} and \ref{fig_bseq}). It is conveniently rewritten as 
\begin{align}
\dk \Phi_k[\calG] ={}& \frac{1}{3!} \tdk \Tr \ln \bigl(\Pi^{-1}+\Phi_k^{(2)}+R_k \bigr)  \nonumber \\ 
& - \frac{1}{12} \sum_{\gam_1,\gam_2} \dot R_{k,\gam_1\gam_2} \calG_{\gam_1} \calG_{\gam_2} ,
\label{rgeq}
\end{align} 
where we have introduced the operator $\tdk=(\dk R_k)\partial_{R_k}$ acting on the $k$ dependence of $R_k$ (but not on that of $\Phi_k^{(2)}$). This exact RG equation leads to an infinite hierarchy of equations for the 2PI vertices $\Phi_k^{(n)}$. In the following we discuss the one- and two-particle vertices.

\subsection{One-particle 2PI vertex}
\label{subsec_dphi1}

Taking the functional derivative of Eq.~(\ref{rgeq}), we obtain a RG equation for the one-particle vertex (or self-energy),\cite{note13}
\begin{align}
\dk \Phi_{k,\gam_1}^{(1)} ={}& \third \tdk \sum_{\gam_2} \calG_{\gam_2} \Delta \calY_{k;\alpha_1\alpha_2,\alpha'_2\alpha_1'} \nonumber \\ & 
+ \frac{1}{3!} \tdk \Tr \left[ W_k^{(2)} \Phi_{k,\gam_1}^{(3)} \right]   
\label{dphi1}
\end{align}
(see Appendix~\ref{app_rgeq}), where $\Delta \calY_k = \calY_k-\calX_k$. Equation~(\ref{dphi1}) is shown diagrammatically in Fig.~\ref{fig_rgeq}. In the RG equation of the self-energy $\bar\Sig_{k}=-\bar\Phi^{(1)}_k$, there is an additional term due to the $k$ dependence of the equilibrium propagator $\bar\calG_k$, 
\begin{equation}
\dk \bar\Phi_{k,\gam}^{(1)} = \dk \Phi_{k,\gam}^{(1)}\bigl|_{\bar\calG_k} + \half \sum_{\gam'} \bar\Phi_{k,\gam\gam'}^{(2)} \dk \bar\calG_{k,\gam'} .
\label{dphi1a}
\end{equation} 
Both terms in the rhs of~(\ref{dphi1}) describe a fermion interacting with a collective (bosonic) fluctuation. The first one is often considered within RPA-like theories; note however that here it includes ``vertex'' corrections as the 2PI vertex $\Phi_k^{(2)}$ implicit in $\Delta\calY_k$ is $k$ dependent. The second one is a purely bosonic term as it involves the correlation function $W_k^{(2)}$. 

\begin{figure}
\centerline{\includegraphics[width=8.cm]{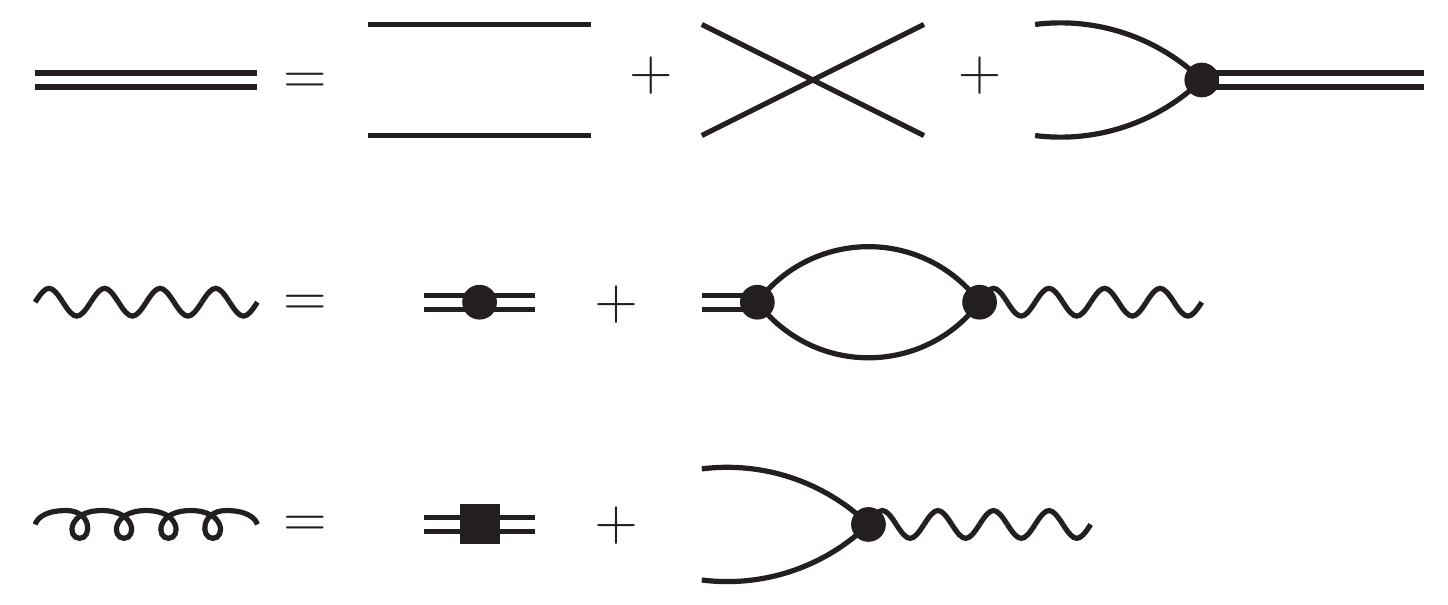}}
\caption{Diagrammatic representation of the Bethe-Salpeter equations satisfied by $W^{(2)}_k$, $\calY_k$ and $\calJ_k$ [Eqs.~(\ref{W2a},\ref{Ydef},\ref{Jdef})]. (Signs and symmetry factors are not shown.)}
\label{fig_bseq}
\end{figure}

Equation~(\ref{dphi1a}) can be projected onto the various channels. As an example, consider a U(1) and spin-rotation invariant system; the only nonzero component is $\bar\Phi_{k,\rm ch}^{(1)}\equiv\bar\Phi_{k,+-}^{(1)0}$. Neglecting the three-particle 2PI vertex $\bar\Phi_k^{(3)}$, we find 
\begin{multline}
\dk \bar\Phi_{k,\rm ch}^{(1)}(x_1,x'_1) = - \frac{1}{6} \tdk \int dx_2dx'_2 \bigl\lbrace \bGchk(x_2,x'_2) \\ \times [ \Delta \bar \calY_{k,\rm ch}(x_1,x_2;x'_2,x'_1)  + 3  \Delta \bar \calY_{k,\rm sp}(x_1,x_2;x'_2,x'_1) ] \\
- \bGchk(x'_2,x_2) [ \Delta \bar \calY_{k,\rm s}(x_1,x_2;x'_2,x'_1) \\ + 3  \Delta \bar\calY_{k,\rm t}(x_1,x_2;x'_2,x'_1) ] \bigr\rbrace \\ 
- \int dx_2dx'_2 \bar\Phi^{(2)}_{k,\rm ch}(x_1,x'_1;x_2,x'_2) \dk \bGchk(x'_2,x_2) , 
\label{dphi1b}
\end{multline} 
where $\Delta \bar \calY_{k,\rm ch}= \Delta \bar \calY_{k,\rm ph}^{00}$, $\Delta \bar \calY_{k,\rm sp}= \Delta \bar \calY_{k,\rm ph}^{\nu\nu}$,  $\Delta \bar \calY_{k,\rm s}= \Delta \bar \calY_{k,\rm pp}^{00}$ and $\Delta \bar \calY_{k,\rm t}= \Delta \bar \calY_{k,\rm pp}^{\nu\nu}$ ($\nu=x,y,z$). Equation~(\ref{dphi1b}) describes the interaction of a fermion with collective ph and pp fluctuations. 

\subsection{Two-particle 2PI vertex}

\begin{figure}
\centerline{\includegraphics[width=8.4cm]{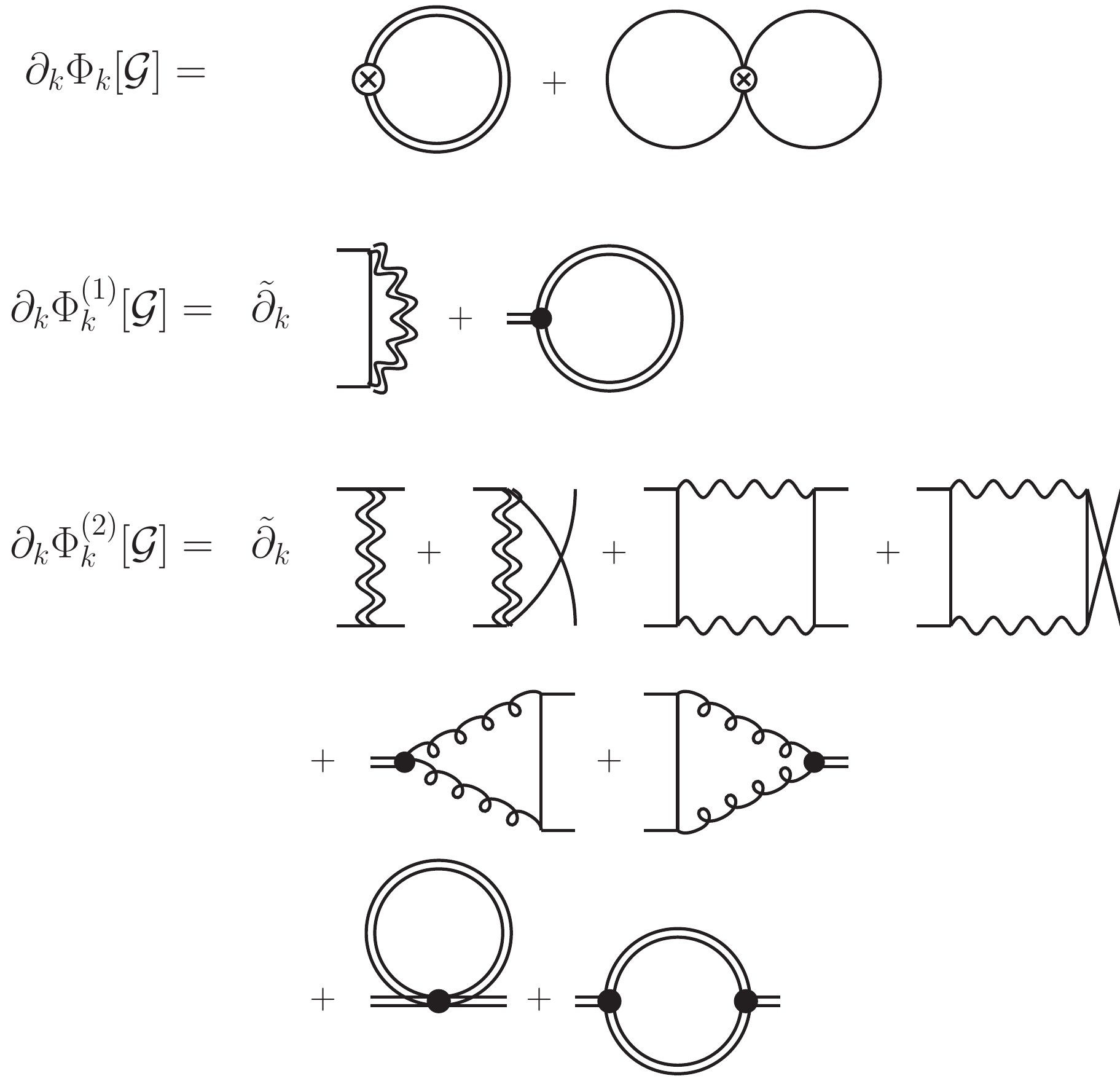}}
\caption{Diagrammatic representation of the RG equations $\dk\Phi_k[\calG]$, $\dk\Phi_k^{(1)}[\calG]$ and $\dk\Phi_k^{(2)}[\calG]$. The double wavy line stands for $\Delta\calY_k$. (Signs and symmetry factors are not shown.)}
\label{fig_rgeq} 
\end{figure}

The second-order functional derivative of Eq.~(\ref{rgeq}) gives 
\begin{multline}
\dk \Phi_{k,\gam_1\gam_2}^{(2)} = \third \tdk [\Delta \calY_{k;\alpha_1\alpha_2,\alpha'_2\alpha_1'} - (\alpha_2 \leftrightarrow \alpha'_2) ]  \\ 
- \third \tdk \sum_{\gam_3,\gam_4} \calG_{\gam_3} \calG_{\gam_4} [ \calY_{k;\alpha_1\alpha_3,\alpha_2\alpha_4} \calY_{k;\alpha'_4\alpha'_2,\alpha'_3\alpha'_1} - (\alpha_2 \leftrightarrow \alpha'_2) ] \\ 
+ \third \tdk \sum_{\gam_3} \left[ \calG_{\gam_3} (\calJ_k^T \Phi^{(3)}_{k,\gam_1} \calJ_k )_{\alpha_2\alpha_3,\alpha_3'\alpha'_2} + (\gam_1 \leftrightarrow \gam_2) \right]  \\
+ \frac{1}{3!} \tdk \Tr \left[ W_k^{(2)} \Phi_{k,\gam_1\gam_2}^{(4)} - \Phi_{k,\gam_1}^{(3)} W_k^{(2)} \Phi_{k,\gam_2}^{(3)}  W_k^{(2)} \right] 
\label{dphi2}   
\end{multline} 
(see Appendix~\ref{app_rgeq} and Fig.~\ref{fig_rgeq}), where 
\begin{equation}
\begin{split}
\calJ_k &= (\calI+\Pi \calX_k)^{-1} = \calI - \Pi \calX_k \calJ_k , \\ 
\calJ^T_k &= (\calI+\calX_k\Pi)^{-1} = \calI - \calX_k \Pi \calJ^T_k . 
\end{split}
\label{Jdef}
\end{equation}
As in the case of $\dk\Phi_k^{(1)}$, there are purely ``bosonic'' terms (involving $W_k^{(2)}$), while other terms clearly exhibit the fermionic nature of the fundamental degrees of freedom. 
The first term in the rhs of~(\ref{dphi2}) has a simple physical interpretation: it expresses the fact that the 1PI vertex in a given channel ``feeds'' the 2PI vertex in other channels. For example, the 1PI vertex $\Delta \calY_{k,\rm sp}^{\nu\nu'}$ in the spin ph channel couples to the 2PI vertex $\Phi_{k,\rm pp}^{(2)\nu\nu'}$ in the pp channel. This coupling is responsible, in the Hubbard model near half-filling, of superconductivity induced by spin fluctuations.\cite{[{See, for instance, }]Scalapino95,*Scalapino12} The terms $\calG\calG \calY_k\calY_k$ are the one-loop terms (without the two-particle reducible contribution) well-known from the 1PI RG approach to fermion systems.\cite{Metzner12} They also contribute to the coupling between channels. The last term in the rhs of~(\ref{dphi2}) describes interactions between collective (bosonic) fluctuations. The two-particle 2PI vertex in the equilibrium state satisfies the equation 
\begin{equation}
\dk \bar\Phi_{k,\gam_1\gam_2}^{(2)} = \dk \Phi_{k,\gam_1\gam_2}^{(2)}\bigl|_{\bar\calG_k} + \half \sum_{\gam_3} \bar\Phi_{k,\gam_1\gam_2\gam_3}^{(3)} \dk \bar\calG_{k,\gam_3} . 
\label{dphi2a} 
\end{equation}

The flow equations~(\ref{dphi1},\ref{dphi1a}) and (\ref{dphi2},\ref{dphi2a}) involve the 1PI vertex $\bar\calY_k$ and the pair propagator $\bar W^{(2)}_k$. One does not, however, consider RG equations for these quantities. Instead, one deals with the 2PI vertex $\bar\Phi^{(2)}_k$, a quantity which is expected to be much less singular (we further discuss this point in Sec.~\ref{subsec_approxsol}). The price to pay is that, in addition to the RG equations, we have to solve the Bethe-Salpeter equations~(\ref{eqbs1}) and (\ref{eqbs1a}) relating $\bar\calY_k$ and $\bar W^{(2)}_k$ to $\bar\Phi^{(2)}_k$. 

\subsection{Thermodynamic potential} 

The thermodynamic potential
\begin{equation}
\Omega_k = \frac{1}{\beta} \Gamma_k[\bar\calG_k] 
\end{equation}
satisfies the RG equation 
\begin{align}
\dk \Omega_k &= \frac{1}{\beta} \dk\Gamma_k\bigl|_{\bar\calG_k} + \frac{1}{2\beta} \sum_\gam \bar\Gamma^{(1)}_{k,\gam} \dk \bar\calG_{k,\gam} \nonumber \\ &
= - \frac{1}{2\beta} \tr \bigl( \dot \calG^{(0)-1}_k \bar\calG_k \bigr) +
\frac{1}{\beta} \dk\Phi_k\bigl|_{\bar\calG_k} .
\end{align}
The last result is derived using $\bar\Gamma^{(1)}_{k,\gam}=0$ [Eq.~(\ref{Gbardef})] and Eq.~(\ref{rgeq6}). Thus we obtain
\begin{align}
\dk\Omega_k ={}& - \frac{1}{2\beta} \tr \bigl( \dot \calG^{(0)-1}_k \bar\calG_k \bigr) + \frac{1}{3!} \Tr [ \dot R_k \bar W_k^{(2)} ] \nonumber \\ & - \frac{1}{12} \sum_{\gam_1,\gam_2} \dot R_{k,\gam_1\gam_2} \bar\calG_{k,\gam_1} \bar\calG_{k,\gam_2} .
\end{align}

\section{Truncation of the Luttinger-Ward functional} 
\label{sec_truncation} 

Standard approximations in the 2PI effective action formalism are based on truncations of the Luttinger-Ward functional where only a subset of diagrams is considered. In this section, we show that starting from a truncated functional $\Phi_k[\calG]$, the RG equation systematically generates higher-order diagrams. We then propose an approximation scheme for solving the flow equations. 

\subsection{Truncation and generation of higher-order diagrams} 
\label{subsec_truncation} 

\begin{figure}
\centerline{\includegraphics[clip,width=8.5cm]{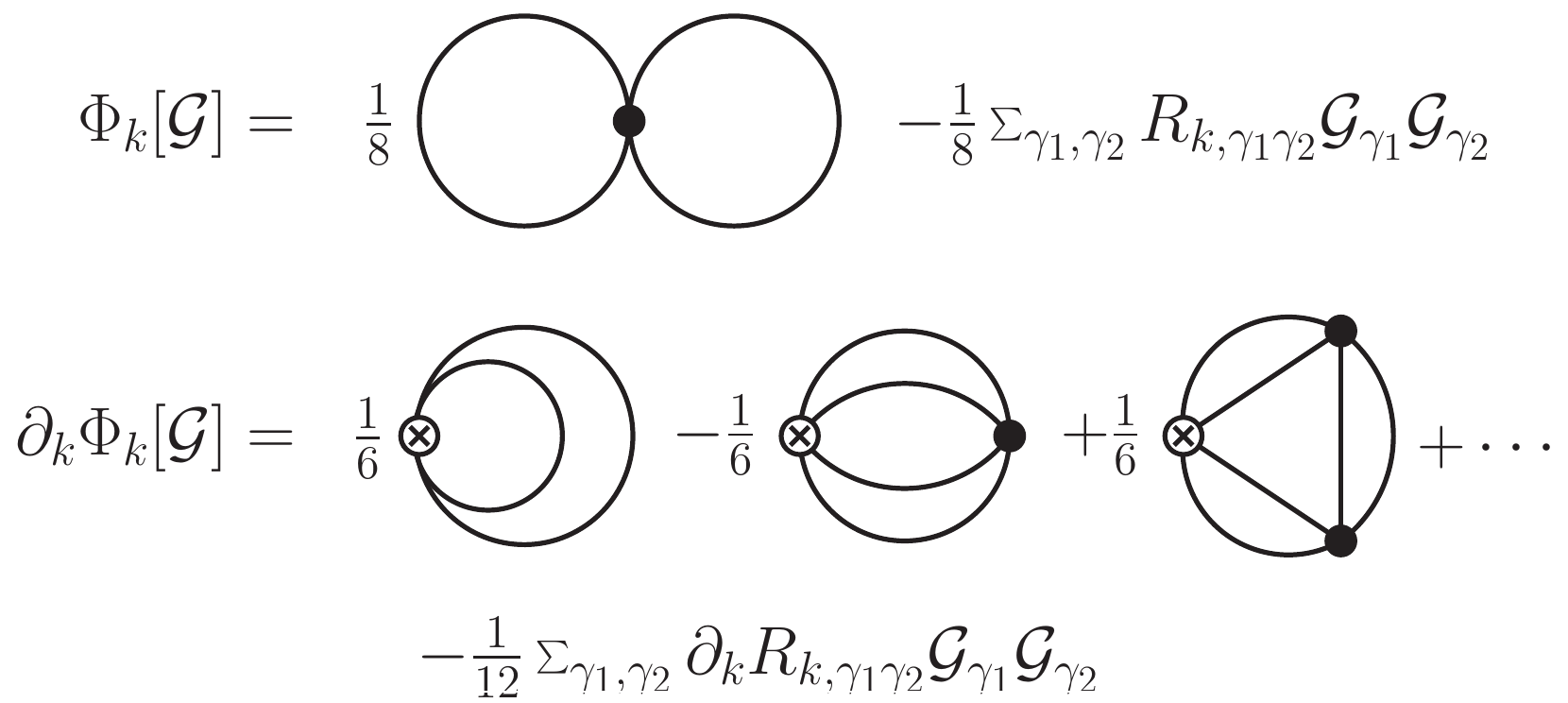}}
\caption{Lowest-order Luttinger-Ward functional  $\Phi_k[\calG]$ [Eq.~(\ref{Phi_2})] and the corresponding RG equation $\dk\Phi_k[\calG]$ [Eq.~(\ref{rgeq2})]. Pairs of solid lines stand for $\Pi$ and the black dot for $U+R_k$.}
\label{fig_lw1} 
\end{figure}

Let us start with the lowest-order $\calO(\calG^2)$ contribution to the Luttinger-Ward functional (Fig.~\ref{fig_lw1}), 
\begin{align}
\Phi_k[\calG] ={}& \frac{1}{8} \sum_{\gam_1,\gam_2} (U+R_k)_{\gam_1\gam_2} \calG_{\gam_1} \calG_{\gam_2} 
- \frac{1}{8} \sum_{\gam_1,\gam_2} R_{k,\gam_1\gam_2} \calG_{\gam_1} \calG_{\gam_2} \nonumber \\ 
={}&  \frac{1}{8} \sum_{\gam_1,\gam_2} U_{\gam_1\gam_2} \calG_{\gam_1} \calG_{\gam_2} ,
\label{Phi_2} 
\end{align}
which is manifestly $k$ independent: $\dk\Phi_k[\calG]=0$. $\dk\Phi_k[\calG]$ can also be computed from the exact RG equation~(\ref{rgeq}). Since $\Phi_k^{(2)}=U$, we find 
\begin{align}
\dk\Phi_k[\calG] ={}& \frac{1}{3!} \Tr[ \dot R_k (\Pi^{-1}+U+R_k)^{-1} ] \nonumber \\ & - \frac{1}{12} \sum_{\gam_1,\gam_2} \dot R_{k,\gam_1\gam_2} \calG_{\gam_1} \calG_{\gam_2} .
\label{rgeq2}
\end{align} 
To $\calO(\calG^2)$, this gives 
\begin{align}
& \frac{1}{3!} \Tr(\dot R_k \Pi) - \frac{1}{12} \sum_{\gam_1,\gam_2} \dot R_{k,\gam_1\gam_2} \calG_{\gam_1} \calG_{\gam_2} \nonumber \\ ={}& \frac{1}{24} \sum_{\gam_1,\gam_2} \dot R_{k,\gam_1\gam_2} \Pi_{\gam_2\gam_1} - \frac{1}{12} \sum_{\gam_1,\gam_2} \dot R_{k,\gam_1\gam_2} \calG_{\gam_1} \calG_{\gam_2} , 
\end{align} 
which vanishes in agreement with the result $\dk\Phi_k[\calG]=0$. However, Eq.~(\ref{rgeq2}) generates higher-order terms which are not included in the original choice of $\Phi_k[\calG]$ (Fig.~\ref{fig_lw1}). The $\calO(\calG^4)$ term gives 
\begin{equation}
- \frac{1}{3!} \Tr[ \dot R_k \Pi (U+R_k)\Pi ] .
\end{equation}
Integrating over $k$ with $R_\Lamb=-U$ and $R_{k=0}=0$, we obtain 
\begin{align}
- \frac{1}{12} \Tr( U \Pi U \Pi ) ={}&  - \frac{1}{48} \sum_{\gam_1\cdots\gam_4} U_{\alpha_3\alpha_4,\alpha_1\alpha_2} U_{\alpha'_1\alpha'_2,\alpha'_3\alpha'_4} \nonumber \\ & \times \calG_{\gam_1} \calG_{\gam_2} \calG_{\gam_3} \calG_{\gam_4} ,
\end{align} 
which is precisely the $\calO(\calG^4)$ contribution to the Luttinger-Ward functional $\Phi_{k=0}[\calG]$. 

\begin{figure}
\centerline{\includegraphics[width=6.5cm]{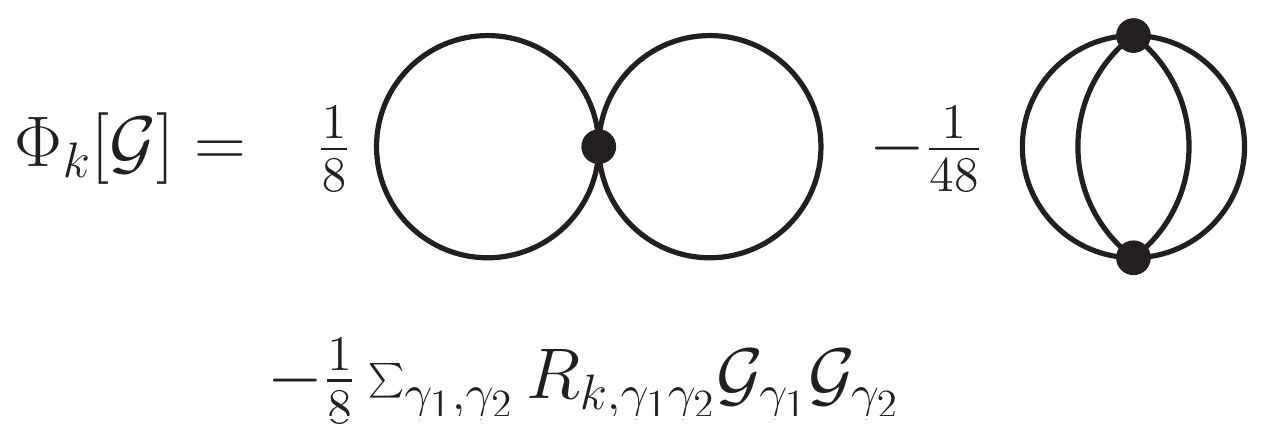}}
\caption{Luttinger-Ward functional $\Phi_k[\calG]$ to $\calO(\calG^4)$.}
\label{fig_lw2}
\end{figure}

Let us now consider the Luttinger-Ward functional with the $\calO(\calG^4)$ contribution included (Fig.~\ref{fig_lw2}), 
\begin{multline}
\Phi_k[\calG] = \frac{1}{8} \sum_{\gam_1,\gam_2} U_{\gam_1\gam_2} \calG_{\gam_1} \calG_{\gam_2} 
- \frac{1}{48} \sum_{\gam_1\cdots\gam_4} (U+R_k)_{\alpha_3\alpha_4,\alpha_1\alpha_2} \\ \times  (U+R_k)_{\alpha'_1\alpha'_2,\alpha'_3\alpha'_4} \calG_{\gam_1} \calG_{\gam_2} \calG_{\gam_3} \calG_{\gam_4} ,
\label{Phi_4} 
\end{multline}
and use again the RG equation to generate higher-order diagrams. Equation~(\ref{Phi_4}) implies 
\begin{multline}
\Phi^{(2)}_{k,\gam_1\gam_2} = U_{\gam_1\gam_2} - \half \sum_{\gam_3,\gam_4} [ (U+R_k)_{\alpha_3\alpha_4,\alpha_1\alpha_2} \\ \times (U+R_k)_{\alpha'_1\alpha'_2,\alpha'_3\alpha'_4} - (\alpha_1 \leftrightarrow \alpha'_1) ] \calG_{\gam_3} \calG_{\gam_4} . 
\end{multline} 
It is straightforward to show that the exact RG equation~(\ref{rgeq}) implies $\dk\Phi_k[\calG]=0$ to $\calO(\calG^2)$. To $\calO(\calG^4)$, we  obtain 
\begin{multline}
- \frac{1}{3!} \Tr[ \dot R_k \Pi (U+R_k)\Pi ] = - \frac{1}{24} \sum_{\gam_1\cdots\gam_4} \dot R_{k,\alpha_3\alpha_4,\alpha_1\alpha_2} \\ \times (U+R_k)_{\alpha'_1\alpha'_2,\alpha'_3\alpha'_4}  \calG_{\gam_1} \calG_{\gam_2} \calG_{\gam_3} \calG_{\gam_4} ,
\end{multline} 
in agreement with the original choice of $\Phi_k[\calG]$ [Eq.~(\ref{Phi_4})]. To next order, $\calO(\calG^6)$, we find
\begin{equation}
\frac{1}{3!} \Tr[\dot R_k \Pi (U+R_k)\Pi (U+R_k)\Pi ] - \frac{1}{3!} \Tr[\dot R_k \Pi(\Phi_k^{(2)}-U) \Pi ] . 
\label{rgeq3} 
\end{equation}
Integrating between $k=\Lamb$ and $k=0$ we obtain 
\begin{align} 
- \frac{1}{18\times 2^3} \sum_{\gam_1\cdots \gam_6} & U_{\alpha'_5\alpha'_6,\alpha_1\alpha_2} U_{\alpha'_1\alpha'_2,\alpha_3\alpha_4} U_{\alpha'_3\alpha'_4,\alpha_5\alpha_6} \nonumber \\ & \times  \calG_{\gam_1} \calG_{\gam_2} \calG_{\gam_3}\calG_{\gam_4} \calG_{\gam_5} \calG_{\gam_6} 
\label{rgeq4} 
\end{align}  
for the first term in~(\ref{rgeq3}), and 
\begin{align}
& - \frac{1}{18\times 2^2} \sum_{\gam_1\cdots \gam_6} U_{\alpha'_5\alpha'_6,\alpha_1\alpha_2} U_{\alpha'_1\alpha'_2,\alpha_3\alpha_4} U_{\alpha'_3\alpha'_4,\alpha_5\alpha_6} \nonumber \\ & \hspace{1.5cm} \times \calG_{\gam_1} \calG_{\gam_2} \calG_{\gam_3}\calG_{\gam_4} \calG_{\gam_5}  \calG_{\gam_6} 
\label{rgeq5} 
\end{align} 
for the second one. Summing~(\ref{rgeq4}) and (\ref{rgeq5}), we recover the $\calO(\calG^6)$ contribution to the Luttinger-Ward functional $\Phi_{k=0}[\calG]$. 

Thus we see that if $\Phi_k[\calG]$ is truncated to a given order, then the exact RG equation $\dk\Phi_k[\calG]$ is correct to that order but generates terms to all (even) orders in $\calG$. The previous calculations suggest that if we start from $\Phi_k[\calG]$ to $\calO(\calG^{2n})$, then $\Phi_{k=0}[\calG]$ obtained from $\dk\Phi_k[\calG]$ is exact to $\calO(\calG^{2n+2})$. 

\subsection{An approximation scheme to solve the RG equations} 
\label{subsec_approxsol}

We propose to approximately solve the RG equations by truncating the Luttinger-Ward functional. However, rather than expanding $\Phi_k[\calG]$ about $\calG=0$ as in the preceding section, we expand about the minimum $\bar\calG_k$ of the 2PI effective action $\Gamma_k[\calG]$. Such a truncation corresponds to a vertex expansion. 

The minimum of the effective action $\Gamma_k[\calG]$ determines the equilibrium state of the system (Sec.~\ref{subsec_eqstate}). If one of the global symmetries of the microscopic action is spontaneously broken, then the minimum is degenerate. Assuming that $\bGspvk$ and $\bFtvk$ are linearly polarized, a particular minimum is defined by the propagators  
\begin{equation}
\begin{gathered}
\bGchk(x,y), \quad 
\bGspk(x,y) \n , \\ 
\bFsk(x,y) e^{2i\theta} = [\bFsk^\dagger(y,x)e^{-2i\theta}]^* , \\ 
\bFtk(x,y) e^{2i\theta} \n' = [\bFtk^\dagger(y,x)e^{-2i\theta}\n'{}^*]^* ,
\end{gathered} 
\label{mink}
\end{equation}
where $\theta$ is an arbitrary phase. $\n$ and $\n'$ are arbitrary real and complex unit vectors, respectively ($\n^2=|\n'|^2=1$). In a normal (nonsuperconducting) phase, $\bar F_{k,\rm s}$ and $\bar F_{k,\rm t}$ vanish; if the system is paramagnetic, $\bar G_{k,\rm sp}$ also vanishes. 

The expansion of $\Phi_k[\calG]$ about the (possibly degenerate) minimum of $\Gamma_k[\calG]$ must respect the global symmetries of the microscopic action~(\ref{action0}), i.e. the translation, SU(2) spin-rotation and U(1) invariances. We must therefore expand $\Phi_k[\calG]$ in terms of the corresponding invariants.\cite{note4} To lowest (quadratic) order, there are four such invariants,\cite{note7} 
\begin{equation}
\begin{gathered}
\Gch(x_1,x'_1) \Gch(x_2,x'_2) , \quad \\ 
\Gspv(x_1,x'_1) \cdot \Gspv(x_2,x'_2) , \\
\Fs^\dagger(x_1,x'_1) \Fs(x_2,x'_2) , \quad \\
\Ftv^\dagger(x_1,x'_1) \cdot \Ftv(x_2,x'_2) 
\end{gathered} 
\label{invdef} 
\end{equation} 
(see also the discussion in Appendix~\ref{app_phi_sym}) and the most general expression of the Luttinger-Ward functional is 
\begin{widetext} 
\begin{align}
\Phi_k[\calG] ={}& \bar\Phi_k 
+ \half \int dx_1 dx'_1 dx_2 dx'_2 \lbrace \ukch(x_1,x'_1;x_2,x'_2) [ \Gch(x'_1,x_1) \Gch(x'_2,x_2) - \bGkch(x'_1,x_1) \bGkch(x'_2,x_2) ]  \nonumber \\ &
+ \uksp(x_1,x'_1;x_2,x'_2) [ \Gspv(x'_1,x_1) \cdot \Gspv(x'_2,x_2) - \bGksp(x'_1,x_1) \bGksp(x'_2,x_2) ] \nonumber \\ &
+ \uks(x_1,x'_1;x_2,x'_2) [ \Fs^\dagger(x_1,x'_1) \Fs(x_2,x'_2) - \bFks^\dagger(x_1,x'_1) \bFks(x_2,x'_2) ] \nonumber \\ &
+ \ukt(x_1,x'_1;x_2,x'_2) [ \Ftv^\dagger(x_1,x'_1) \cdot \Ftv(x_2,x'_2) - \bFkt^\dagger(x_1,x'_1) \bFkt(x_2,x'_2)] \rbrace ,
\label{Phi_5}
\end{align}
\end{widetext}
where $\bar\Phi_k$ denotes the value of $\Phi_k[\calG]$ at the minimum of $\Gamma_k[\calG]$. Note that even when global symmetries are spontaneously broken, the symmetry properties of the Luttinger-Ward functional imply that the 2PI vertex $u_k$ is parameterized only by four independent functions ($\ukch$, etc.). In the quadratic approximation, $\Phi_k[\calG]$ is thus entirely determined by $\bGkch$, $\bGksp$, $\bFks$, $\bFkt$, and $\ukch$, $\uksp$, $\uks$, $\ukt$. 

The 2PI vertex $u_k=\bar\Phi_k^{(2)}$ is obtained from the RG equations~(\ref{dphi2}) and (\ref{dphi2a}) with $\Phi_k^{(3)}=\Phi_k^{(4)}=0$, as well as the Bethe-Salpeter equation relating $\bar\calY_k$ to $u_k$. In many cases, it is possible to ignore part of the momentum-frequency dependence of $u_k$ because that of the 1PI vertex $\bar\calY_k$ mainly comes from the pair propagator $\bar\Pi_k$.\cite{note14} Introducing the total and relative momentum-frequency of the pairs [Eq.~(\ref{qldef1})], we consider the 2PI vertex $u_k(q_1,l_1;q_2,l_2)$ in Fourier space and assume that we can neglect its dependence on the frequency component of $l_1$ and $l_2$. Translation invariance implies $u_k(q_1,\l_1;q_2,\l_2)=\delta_{q_1,q_2} u_k(q_1;\l_1,\l_2)$. We next expand the 2PI vertex
\begin{equation} 
u^{\nu_1\nu_2}_{k,a}(q;\l_1,\l_2) = \sum_{n_1,n_2} u^{\nu_1n_1,\nu_2n_2}_{k,a}(q)  
f_{n_1}(\l_1) f_{n_2}(\l_2) 
\label{Xql} 
\end{equation}
(for a similar expansion of the 1PI vertices, see Ref.~\onlinecite{Husemann09}), 
where $a={\rm pp,ph}$ and the $f_n$'s are form factors satisfying
\begin{equation}
\frac{1}{N} \sum_\l f_n(\l) f_m(\l) = \delta_{n,m} .
\end{equation}
For a square lattice, $f_n(\l)=1$ for $s$-wave, $f_n(\l)=\cos l_x -\cos l_y$ for $d_{x^2-y^2}$-wave, etc. In practice, only a few channels (corresponding to strong fluctuations) are expected to be important. The Bethe-Salpeter equations~(\ref{eqbs1}) and (\ref{eqbs1a}) become 
\begin{multline}  
\calY^{\nu_1 n_1,\nu_2 n_2}_{k,c_1c'_1c_2c'_2}(q_1,q_2) = \calX^{\nu_1 n_1,\nu_2 n_2}_{k,c_1c'_1c_2c'_2}(q_1,q_2)
 \\  - \frac{1}{4} \sum_{c_3\cdots c'_4 \atop \nu_3,\nu_4} \sum_{q_3,q_4 \atop n_3,n_4}  \calX^{\nu_1n_1,\nu_3n_3}_{k,c_1c'_1c_3c'_3}(q_1,q_3) \\ \times \Pi^{\nu_3n_3,\nu_4n_4}_{c_3c'_3c_4c'_4}(q_3,q_4) \calY^{\nu_4n_4,\nu_2n_2}_{k,c_4c'_4c_2c'_2}(q_4,q_2) 
\label{eqbs4}
\end{multline}
and  
\begin{multline}  
W^{(2)\nu_1 n_1,\nu_2 n_2}_{k,c_1c'_1c_2c'_2}(q_1,q_2) = \Pi^{\nu_1 n_1,\nu_2 n_2}_{c_1c'_1c_2c'_2}(q_1,q_2)
 \\  - \frac{1}{4} \sum_{c_3\cdots c'_4 \atop \nu_3,\nu_4} \sum_{q_3,q_4 \atop n_3,n_4}  \Pi^{\nu_1n_1,\nu_3n_3}_{c_1c'_1c_3c'_3}(q_1,q_3) \\ \times \calX^{\nu_3n_3,\nu_4n_4}_{k,c_3c'_3c_4c'_4}(q_3,q_4) W^{(2)\nu_4n_4,\nu_2n_2}_{k,c_4c'_4c_2c'_2}(q_4,q_2) . 
\label{eqbs4a}
\end{multline}
We have introduced 
\begin{align}
\Pi^{\nu_1n_1,\nu_2n_2}_{c_1c'_1c_2c'_2}(q_1,q_2) ={}& \frac{1}{\beta N} \sum_{l_1,l_2} f_{n_1}(\l_1) f_{n_2}(\l_2) \nonumber \\ & \times \Pi^{\nu_1\nu_2}_{c_1c'_1c_2c'_2}(q_1,l_1;q_2,l_2) ,
\label{Pi1} 
\end{align}
and $\calX^{\nu_1 n_1,\nu_2 n_2}_{k,c_1c'_1c_2c'_2}$ and $\calY^{\nu_1 n_1,\nu_2 n_2}_{k,c_1c'_1c_2c'_2}$ are defined as $u^{\nu_1n_1,\nu_2n_2}_{k,c_1c'_1c_2c'_2}$. If the global U(1) symmetry is not broken, Eqs.~(\ref{eqbs4}) and (\ref{eqbs4a}) can be solved independently in the pp and ph channels. A further approximation consists in ignoring the momentum-frequency dependence of $u^{\nu_1n_1,\nu_2n_2}_{k,a}(q)$. The 2PI vertex is then parameterized by a coupling constant $u^{\nu_1n_1,\nu_2n_2}_{k,a}$ for each pair of fluctuation channels $(a,\nu_1,n_1)$ and $(a,\nu_2,n_2)$. The Bethe-Salpeter equations~(\ref{eqbs4}) and (\ref{eqbs4a}) can then easily be solved, in particular when only a small number of channels is taken into account. In this approximation, contrary to the 1PI RG approach, there is no need to discretize the momentum space into patches to keep track of the momentum dependence of the two-particle 1PI vertex $\bar\calY_k$ when solving numerically the flow equations. 

The self-energy $\bar\Sigma_{k,\gam}=-\bar\Phi^{(1)}_{k,\gam}$ can be directly deduced from~(\ref{Phi_5}), 
\begin{equation}
\begin{split} 
\bar\Sig^\nu_{k,\rm ph}(x_1,x'_1) ={}& \int dx_2 dx'_2 u^{\nu\nu}_{k,\rm ph}(x_1,x'_1;x_2,x'_2) \\ & \times 
\bar G^\nu_{k}(x'_2,x_2) , \\
\bar\Sig^\nu_{k,\rm pp}(x_1,x'_1) ={}& - \half \int dx_2 dx'_2 u^{\nu\nu}_{k,\rm pp}(x_1,x'_1;x_2,x'_2) \\ & \times \bar F^\nu_{k}(x_2,x'_2) ,  
\end{split}
\label{self4} 
\end{equation}
where $u^{00}_{k,\rm ph}=\ukch$, $u^{\nu\nu}_{k,\rm ph}=\uksp$, $u^{00}_{k,\rm pp}=\uks$ and $u^{\nu\nu}_{k,\rm pp}=\ukt$ ($\nu\neq 0$). Equations~(\ref{self4}) are similar to the Hartree-Fock approximation but with a ($k$-dependent) momentum-frequency dependent interaction $u_k$. When the minimum $\bar\calG_k$ of $\Gamma_k[\calG]$ is degenerate, the solution of~(\ref{self4}) is not unique. In that case, it is sufficient to choose a particular minimum, compute the corresponding self-energy $\bar\Sig_k$ and deduce $\bGkch$, $\bGksp$, $\bFks$, $\bFkt$ using~(\ref{mink}). 

The discussion of Sec.~\ref{subsec_truncation} suggests that it may be advantageous to determine the $k$-dependent self-energy $\bar\Sig_k=-\bar\Phi_k^{(1)}$ from its RG equation~(\ref{dphi1a}), 
\begin{equation}
\dk \bar\Sig_{k,\gam} = - \dk \Phi_{k,\gam}^{(1)}\bigl|_{\bar\calG_k} - \half \sum_{\gam'} u_{k,\gam\gam'} \dk \bar\calG_{k,\gam'} ,
\label{self5} 
\end{equation}
rather than directly from the Luttinger-Ward functional~(\ref{Phi_5}). Given the RG equation of the vertex $\Phi^{(1)}_k$ [Eq.~(\ref{dphi1})], the interaction of fermions with collective fluctuations is made explicit in~(\ref{self5}). In Eq.~(\ref{self4}), this interaction is hidden in the momentum-frequency dependence of the 2PI vertex $u_k$. Since such a dependence is difficult to take into account in the numerical solution of the flow equations,\cite{note9} Eq.~(\ref{self4}) is of little use in practice. By contrast, Eq.~(\ref{self5}) always provides a nontrivial momentum-frequency dependence of the self-energy $\bar\Sig_k$ even when the 2PI vertex $u_k$ is approximated by a set of coupling constants $\lbrace u^{\nu_1n_1,\nu_2n_2}_{k,a}\rbrace$ as discussed above.  

We can ask whether the quadratic ansatz~(\ref{Phi_5}) with momentum-frequency independent 2PI vertices $u^{\nu_1n_1,\nu_2n_2}_{k,a}$ is justified in the strong-coupling limit. Since such an ansatz encompasses the Hartree-Fock--RPA theory, it captures at least some of the strong-coupling effects in the large-$U$ Hubbard model (see Sec.~\ref{subsec_init}) but there is no guarantee that it is always sufficient. There is no conceptual difficulty in considering momentum-frequency dependent vertices $u^{\nu_1n_1,\nu_2n_2}_{k,a}(q)$ but this of course will make the numerical treatment of the flow equations slightly more difficult. Including higher-order 2PI vertices is also possible. For instance, in a system with strong antiferromagnetic fluctuations, it might be necessary to include a quartic term of the form 
\begin{equation}
\frac{v_{k,\rm sp}}{8} \int dx \left[ \Gspv(x,x^+)^2 - \bGspk(x,x^+)^2 \right]^2 
\end{equation}
in the Luttinger-Ward functional. Whether such improvements are necessary or not is an open question.  

A last comment regards the fulfillment of the Mermin-Wagner theorem. If the equilibrium state spontaneously breaks a continuous symmetry, collective Goldstone modes show up in the two-particle vertex $\bar\calY_k$ and the pair propagator $\bar W^{(2)}_k$. In the 2PI-NPRG approach, these modes are regularized in the infrared by the cutoff function $R_k$, and they appear as poles of the form $A_k(c^2_k\q^2+\wnu^2)+R_k(q)$ (Sec.~\ref{subsec_goldstone}). As a result the self-energy equation $\dk \bar \Sig_k(\p,i\wn)$ receives contributions of the form 
\begin{equation}
\tdk T \sum_{\wnu} \int \frac{d^dq}{(2\pi)^d} \frac{\bar G_k(\p+\q,i\wn+i\wnu)}{A_k(c_k^2\q^2+\wnu^2)+R_k(\q)}  ,
\label{self6}  
\end{equation}
where $\wn$ and $\wnu$ are fermionic and bosonic Matsubara frequencies, respectively. Here we assume that $R_k(q)$ depends only on $\q$ (which is likely to be the case in practice). At finite temperature and in two dimensions, the momentum integral for $\wnu=0$ in~(\ref{self6}) is convergent only if $R_k(\q)$ is nonzero. In this case, when $k\to 0$ and $R_k(\q)\to 0$, the broken symmetry must necessarily been restored, in agreement with the Mermin-Wagner theorem.\cite{note5}

\subsection{Cutoff function $R_k$} 
\label{subsec_R} 

In this section, we propose a simple expression for the cutoff function assuming the lattice to be hypercubic. $R_k$ can be decomposed onto the various fluctuation channels as the 2PI vertex $\calX_k$ (Sec.~\ref{subsec_approxsol}). It is therefore defined by its components $R^{\nu_1n_1,\nu_2n_2}_{k,c_1c'_1c_2c'_2}(q_1;q_2)$, where $q$ denotes the total momentum-frequency of the pair and the indices $n_1,n_2$ refer to the form factors $f_n$. Since the regulator term~(\ref{DSk}) must be translation and spin-rotation invariant, we must have $\nu_1=\nu_2$ and $q_1=q_2$. We choose a cutoff function which is diagonal in the index $n$,
\begin{equation}
R^{\nu_1 n_1,\nu_2 n_2}_{k,c_1c'_1c_2c'_2}(q_1,q_2) = \delta_{q_1,q_2} \delta_{\nu_1,\nu_2} \delta_{n_1,n_2} R^{\nu_1 n_1}_{k,c_1c'_1c_2c'_2}(q_1) .
\end{equation}
Furthermore, since $\Delta S_k$ must satisfy the global U(1) invariance, we can restrict ourselves to 
\begin{equation}
\begin{split}
R^{\nu n}_{k,\rm pp}(q) &= R^{\nu n}_{k,++--}(q) , \\ 
R^{\nu n}_{k,\rm ph}(q) &= R^{\nu n}_{k,+-+-}(q) .
\end{split}
\end{equation}

\begin{figure}
\centerline{\includegraphics[width=5.5cm]{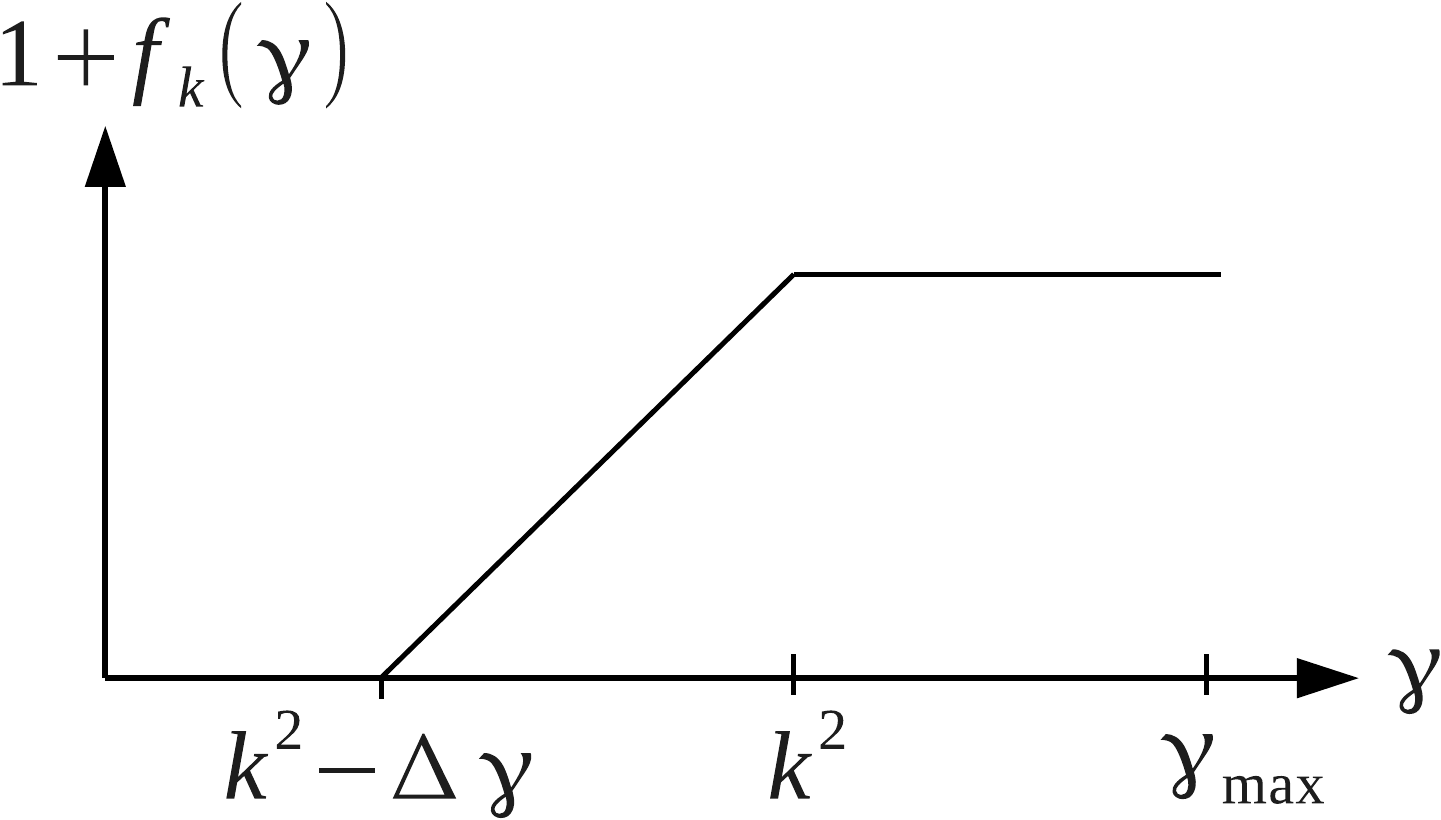}}
\caption{$1+f_k(\gam)$ vs $\gam$ for a typical value of $k$ ($\Delta\gam<k^2<\gam_{\rm max}$) [Eq.~(\ref{fkdef})].}
\label{fig_fk}
\end{figure}

Let us consider a particular channel $(a,\nu,n)$ and denote by $u^{\nu n}_{k,a}(q)\equiv u^{\nu n,\nu n}_{k,a}(q,q)$ the corresponding two-particle 2PI vertex obtained from the quadratic ansatz~(\ref{Phi_5}). We choose a cutoff function of the form 
\begin{equation}
R^{\nu n}_{k,a}(q) = u^{\nu n}_{k,a}(q) f_k(\gam_{\q-\Q}) , 
\end{equation}
where 
\begin{equation}
\gam_\q = 2d - 2 \sum_{i=1}^d \cos q_i .
\end{equation}
A possible choice for the function $f_k$ is 
\begin{equation}
f_k(\gam) = -1 + \Theta(\Delta\gam-k^2+\gam) \Bigl[ 1 + \frac{\gam-k^2}{\Delta\gam} \Theta(k^2-\gam) \Bigr]. 
\label{fkdef}
\end{equation}
Here $\Q$ denotes the momentum where fluctuations in the considered channel are most important (e.g. $\Q=(\pi,\pi)$ for the ($s$-wave) spin ph channel in the two-dimensional Hubbard model at half-filling), and $\Delta\gam$ is an adjustable parameter. The maximum (i.e. initial) value of $k$ is $\Lamb=(\gam_{\rm max}+\Delta\gam)^{1/2}$ with $\gam_{\rm max}=4d$. $f_{\Lamb}(\gam)=-1$ and $f_{k=0}(\gam)=0$, which implies $R_{\Lamb}=-U$ and $R_{k=0}=0$ as it should. 

Since $R_k$ always enters the propagators in the combination $\calX_k=R_k+\Phi_k^{(2)}$, what matters is 
\begin{equation}
R^{\nu n}_{k,a}(q) + u^{\nu n}_{k,a}(q) = u^{\nu n}_{k,a}(q) [ 1 + f_k(\gam_{\q-\Q}) ] .
\end{equation}
Figure~\ref{fig_fk} clearly shows that $R_k$ cancels the fermion-fermion interaction for $\gam_{\q-\Q}< k^2-\Delta\gam$. In other words, $R_k$ tends to suppress low-energy bosonic fluctuations (with a total pair momentum such that $\gam_{\q-\Q}<k^2$) while leaving high-energy fluctuations unchanged.

\section{Summary and conclusion} 
\label{sec_conclu}

We have discussed a NPRG approach in the 2PI effective action formalism which differs in many respects from the more standard 1PI formalism.\cite{Metzner12} The scale-dependent 2PI effective action $\Gamma_k[\calG]$ and Luttinger-Ward functional  $\Phi_k[\calG]$ are constructed by introducing in the action both a (quadratic) fermionic and a (quartic) bosonic regulator. The exact RG equation satisfied by $\Phi_k[\calG]$ is determined by the bosonic cutoff function $R_k$ (the fermionic cutoff function $R_k^{(F)}$ playing only a secondary role) and appears as a bosonic analog of the RG equation satisfied by the 1PI effective action in the 1PI formalism.\cite{Wetterich93,Berges02,Delamotte12,Kopietz_book} In this respect, the 2PI-NPRG is reminiscent of the partial bosonization approach where bosonic fields are introduced at a very early stage {\it via} a Hubbard-Stratonovich transformation of the interaction term in the action.\cite{Baier00,Baier04,Baier05,Schutz05,Krahl07,Krahl09a,Friederich11} Both approaches 
emphasize the 
importance of collective (bosonic) fluctuations and the necessity to control them. 

The control of the bosonic fluctuations and the absence of divergence of two-particle vertices and correlation functions for $k>0$ allows us to describe phases with spontaneously broken symmetries. In particular, we find that the bosonic regulator introduces a mass in the Goldstone mode propagator, which vanishes only for $k\to 0$. The initial condition of the flow is the Hartree-Fock--RPA theory (possibly including spontaneously broken symmetries) where the one-particle propagator includes the fermionic regulator. In the absence of the latter, we recover the standard Hartree-Fock--RPA theory. We have discussed in detail the initial condition of the flow for the two-dimensional half-filled Hubbard model in the large-$U$ limit and shown (in the absence of the fermionic regulator) that it reproduces the spin-wave mode spectrum obtained from the Heisenberg model (in the spin-wave approximation) with exchange coupling constant $J=4t^2/U$.\cite{Schrieffer89,Singh90,
Chubukov92,Borejsza04} This indicates that the 2PI-NPRG captures, already at the level of the initial condition of the flow, some aspects of the strong-coupling limit of the Hubbard model. 

In the 2PI-NPRG approach, the RG equations involve not only the 2PI vertices but also the two-particle 1PI vertex $\bar\calY_k$ and the pair propagator $\bar W^{(2)}_k$ in the equilibrium state. Thus it is necessary, in addition to the RG equations, to solve the Bethe-Salpeter equations relating $\bar\calY_k$ and $\bar W^{(2)}_k$ to $\bar\Phi^{(2)}_k$. On the other hand, the 2PI vertices are much less singular than their 1PI counterparts.\cite{Dupuis05} We therefore expect simple approximations, where the 2PI vertices $\bar\Phi_k^{(n)}$ ($n\geq 2$) are parameterized by a few coupling constants, to be reliable. (As far as possible, one would like to preserve the full momentum-frequency dependence of the self-energy $\bar\Sig_k=-\bar\Phi^{(1)}_k$.) This reduces the functional flow equations for the $\bar\Phi_k^{(n)}$'s ($n\geq 2$) to a finite number of equations for coupling constants, and also simplifies the solution of the Bethe-Salpeter equations. 

In Sec.~\ref{subsec_approxsol} we have proposed an approximation scheme to solve the RG equations where the Luttinger-Ward functional $\Phi_k[\calG]$ is expanded to quadratic order about the minimum $\bar\calG_k$ of the 2PI effective action $\Gamma_k[\calG]$ while satisfying the global symmetries of the action. If we further approximate the two-particle 2PI vertex by a small number of coupling constants, we end up with RG equations which can be solved at modest numerical cost. Work in that direction will be reported in a future publication. 

Among various possible applications of the 2PI-NPRG, we would like to mention systems with strong collective fluctuations, a situation where the weak-coupling 1PI fermionic RG becomes uncontrolled. In particular the RG equation satisfied by the self-energy (which makes the coupling between fermions and collective fluctuations apparent) should allow us to better understand the behavior of a fermion system in the vicinity of a quantum phase transition (e.g. the transition between an antiferromagnet and a metal). 

In a recent paper, Kemler and Braun have discussed a RG approach in the 2PI formalism in the context of density functional theory (DFT).\cite{Kemler13} In their approach, the external source $J_\sig(\r,\tau)$ couples to the density field $\psi^*_\sig(\r,\tau)\psi_\sig(\r,\tau)$ and the 2PI effective action $\Gamma_k[\rho]$ is a functional of the classical variable $\rho_\sig(\r,\tau)=\mean{\psi^*_\sig(\r,\tau)\psi_\sig(\r,\tau)}$. While less general than the effective action $\Gamma_k[\calG]$ that we have discussed, the functional $\Gamma_k[\rho]$ allows us to make a direct connection with DFT and might provide us with a powerful tool to compute the Hohenberg-Kohn functional from a microscopic model. 

The 2PI-NPRG approach can also be used to derive a functional $\Gamma_k[n]$ of the Wigner distribution function $n\equiv\lbrace n_{\p\sig}(\r,\tau)\rbrace$.\cite{note16} In Ref.~\onlinecite{Dupuis00}, it was shown that this functional allows us to make the connection with Fermi-liquid theory and derive the quantum Boltzmann equation satisfied by $n$.

\begin{acknowledgments}
We thank T. Debelhoir for a critical reading of the manuscript. 
\end{acknowledgments}

\appendix

\section{Bosonic matrix formalism}
\label{app_bform} 

Because of the anticommutation relations satisfied by the Grassmann variables $\psi_\alpha$, the one-particle propagator $\calG_\gam$ and the external source $J_\gam$ are antisymmetric under the exchange $\alpha\leftrightarrow\alpha'$ (with $\gam=\lbrace\alpha,\alpha'\rbrace$). We thus define the ``bosonic'' unit matrix
\begin{equation}
\calI_{\gam_1\gam_2} = \frac{\delta J_{\gam_1}}{\delta J_{\gam_2}} = \delta_{\alpha_1,\alpha_2} \delta_{\alpha'_1,\alpha'_2} - \delta_{\alpha_1,\alpha'_2} \delta_{\alpha'_1,\alpha_2} .
\label{Idef} 
\end{equation} 
For any bosonic matrix $A_{\gam_1\gam_2}$ which is antisymmetric under the exchange $\alpha_1\leftrightarrow\alpha_1'$ or $\alpha_2\leftrightarrow\alpha_2'$, we define the trace and the inverse matrix by  
\begin{equation}
\Tr A = \half \sum_\gam A_{\gam\gam} , \qquad A A^{-1} = A^{-1} A = \calI . 
\label{Trdef}
\end{equation}
The product of two matrices $A$ and $B$ is defined by 
\begin{equation}
(AB)_{\gam_1\gam_2} = \half \sum_{\gam_3} A_{\gam_1\gam_3} B_{\gam_3\gam_2} .
\end{equation}
The antisymmetry under the exchange $\alpha\leftrightarrow\alpha'$ also implies that the chain rule for derivation is defined with an additional factor $1/2$, e.g. 
\begin{equation}
\frac{\delta W[J]}{\delta \calG_\gam} = \half \sum_{\gam'} \frac{\delta W[J]}{\delta J_{\gam'}}
\frac{\delta J_{\gam'}}{\delta \calG_\gam} . 
\label{chainrule} 
\end{equation}
The equation of motion~(\ref{eos}) follows from Eq.~(\ref{chainrule}).

\section{Fourier transforms} 
\label{app_fourier}

In this Appendix, we summarize the definitions of the Fourier transforms. To alleviate the notations, we drop the $k$ index in Appendices~\ref{app_fourier} and \ref{app_bseq}. 

\subsection{Fields}

\begin{equation}
\begin{split}
\psi_{c\sig}(x) &= \frac{1}{\sqrt{\beta N}} \sum_p e^{-icpx} \psi_{c\sig}(p) , \\ 
\psi_{c\sig}(p) &= \frac{1}{\sqrt{\beta N}} \int dx\, e^{icpx} \psi_{c\sig}(x) ,
\end{split}
\end{equation}
where $p=(\p,i\wn)$ and $px=\p\cdot\r-\wn\tau$. 

\subsection{Propagators}

\begin{equation} 
\begin{split}
\calG^\nu_{cc'}(x,x') &= \frac{1}{\beta N} \sum_{p,p'} e^{-i(cpx+c'p'x')} \calG^\nu_{cc'}(p,p') , \\
\calG^\nu_{cc'}(p,p') &= \frac{1}{\beta N} \int dx\,dx'\, e^{i(cpx+c'p'x')} \calG^\nu_{cc'}(x,x') , 
\end{split}
\end{equation}
\begin{widetext}
\begin{equation}
\begin{split}
\Pi_{c_1c_1'c_2c_2'}^{\nu_1\nu_2}(x_1,x_1';x_2,x_2')  &= \frac{1}{(\beta N)^2} \sum_{p_1\cdots p_2'} \, e^{-i(c_1p_1x_1+c_1'p_1'x_1'+c_2p_2x_2+c_2'p_2'x_2')}  \Pi_{c_1c_1'c_2c_2'}^{\nu_1\nu_2}(p_1,p_1';p_2,p_2') , \\ 
\Pi_{c_1c_1'c_2c_2'}^{\nu_1\nu_2}(p_1,p_1';p_2,p_2') &= \frac{1}{(\beta N)^2} \int dx_1dx'_1dx_2 dx_2' \, e^{i(c_1p_1x_1+c_1'p_1'x_1'+c_2p_2x_2+c_2'p_2'x_2')} \Pi_{c_1c_1'c_2c_2'}^{\nu_1\nu_2}(x_1,x_1';x_2,x_2') 
\end{split}
\end{equation}
(and similarly for $W^{(2)}$).

\subsection{Vertices}

\begin{equation} 
\begin{split}
Y_{c_1c_1'c_2c_2'}^{\nu_1\nu_2}(x_1,x_1';x_2,x_2')  &= \frac{1}{(\beta N)^3} \sum_{p_1\cdots p_2'} \, e^{i(c_1p_1x_1+c_1'p_1'x_1'+c_2p_2x_2+c_2'p_2'x_2')}  Y_{c_1c_1'c_2c_2'}^{\nu_1\nu_2}(p_1,p_1';p_2,p_2') , \\ 
Y_{c_1c_1'c_2c_2'}^{\nu_1\nu_2}(p_1,p_1';p_2,p_2') &= \frac{1}{\beta N} \int dx_1dx'_1dx_2 dx_2' \, e^{-i(c_1p_1x_1+c_1'p_1'x_1'+c_2p_2x_2+c_2'p_2'x_2')} Y_{c_1c_1'c_2c_2'}^{\nu_1\nu_2}(x_1,x_1';x_2,x_2') 
\end{split}
\end{equation}
(and similarly for $\calX$ and $\Phi^{(2)}$).
\end{widetext}

\section{Propagators, vertices and Bethe-Salpeter equations}
\label{app_bseq}

In this Appendix, we review basic properties of the propagators and vertices and derive the Bethe-Salpeter equations~(\ref{eqbs1}) and (\ref{eqbs1a}).

\subsection{Propagators} 

The pair propagator $W_k^{(2)}$ satisfies
\begin{equation}
\begin{split}
& W^{(2)}_{c_1c_1'c_2c_2'}(X_1,X_1';X_2,X_2') \\  
={}& \quarter \sum_{\nu_1,\nu_2} \bigl(\tau^{\nu_1}_{c_1c_1'}\bigr)_{\sig_1\sig_1'} \bigl(\tau^{\nu_2}_{c_2c_2'}\bigr)_{\sig_2\sig_2'} W^{(2)\nu_1\nu_2}_{c_1c_1'c_2c_2'}(x_1,x'_1;x_2,x'_2), \\
& W^{(2)\nu_1\nu_2}_{c_1c_1'c_2c_2'}(x_1,x'_1;x_2,x'_2) \\ 
={}& \sum_{\sig_1\cdots\sig_2'} \bigl(\tau^{\nu_1\dagger}_{c_1c_1'}\bigr)_{\sig_1'\sig_1} \bigl(\tau^{\nu_2\dagger}_{c_2c_2'}\bigr)_{\sig_2'\sig_2} W^{(2)}_{c_1c_1'c_2c_2'}(X_1,X_1';X_2,X_2') .
\end{split}
\end{equation}   
In Fourier space, it is convenient to introduce the total and relative momentum-frequency of the pair, 
\begin{equation}
W_{c_1c_1'c_2c_2'}^{(2)\nu_1\nu_2}(p_1,p_1';p_2,p_2') = W_{c_1c_1'c_2c_2'}^{(2)\nu_1\nu_2}(q_1,l_1;q_2,l_2) ,
\end{equation}
where
\begin{equation}
\llbrace 
\begin{array}{l}
q_1 = p_1'+c_1c_1' p_1 , \\ 
l_1 = \half(p_1'-c_1c_1' p_1) , 
\end{array}
\right. \quad \mbox{and} \quad  
\llbrace 
\begin{array}{l}
q_2 = p_2+c_2c_2' p_2' , \\ 
l_2 = \half(p_2-c_2c_2' p_2')  
\end{array}
\right.
\label{appbs1}
\end{equation} 
(similar expressions hold for $\Pi$).

\subsection{Vertices} 

The two-particle 2PI vertex satisfies
\begin{equation}
\begin{split}
& \Phi^{(2)}_{c_1c_1'c_2c_2'}(X_1,X_1';X_2,X_2') \\ ={}& \sum_{\nu_1,\nu_2}  \bigl(\tau^{\nu_1\dagger}_{c_1c_1'}\bigr)_{\sig_1'\sig_1} \bigl(\tau^{\nu_2\dagger}_{c_2c_2'}\bigr)_{\sig_2'\sig_2} \Phi^{(2)\nu_1\nu_2}_{c_1c_1'c_2c_2'}(x_1,x_1';x_2,x_2') , \\ 
& \Phi^{(2)\nu_1\nu_2}_{c_1c_1'c_2c_2'}(x_1,x_1';x_2,x_2') \\ ={}& \quarter \sum_{\sig_1\cdots \sig_2'} \bigl(\tau^{\nu_1}_{c_1c_1'}\bigr)_{\sig_1\sig_1'} \bigl(\tau^{\nu_2}_{c_2c_2'}\bigr)_{\sig_2\sig_2'} 
\Phi^{(2)}_{c_1c_1'c_2c_2'}(X_1,X_1';X_2,X_2') 
\end{split}
\label{phi2nu}
\end{equation}
and 
\begin{equation}
\Phi^{(2)\nu_1\nu_2}_{c_1c_1'c_2c_2'}(p_1,p_1';p_2,p_2') = \Phi^{(2)\nu_1\nu_2}_{c_1c_1'c_2c_2'}(q_1,l_1;q_2,l_2) ,
\end{equation}
where
\begin{equation}
\llbrace 
\begin{array}{l}
q_1 = p_1+c_1c_1' p_1' , \\ 
l_1 = \half(p_1-c_1c_1' p_1') ,
\end{array}
\right.\quad \mbox{and} \quad  
\llbrace 
\begin{array}{l}
q_2 = p_2'+c_2c_2' p_2 , \\ 
l_2 = \half(p_2'-c_2c_2' p_2)   
\end{array}
\right.
\label{appbs2}
\end{equation}
(similar expressions hold for $\calX$ and $\calY$). Note that~(\ref{appbs2}) slightly differs from~(\ref{appbs1}). 

\subsection{Bethe-Salpeter equations}

The Bethe-Salpeter equation $\calY=\calX-\calX\Pi\calY$ reads 
\begin{multline}
\calY_{c_1c'_1c_2c'_2}(X_1,X'_1;X_2,X'_2) = \calX_{c_1c'_1c_2c'_2}(X_1,X'_1;X_2,X'_2) \\ 
 - \quarter \sum_{c_3\cdots c'_4} \int dX_3dX'_3 dX_4 dX'_4 \, \calX_{c_1c'_1c_3c'_3}(X_1,X'_1;X_3,X'_3)  \\ \times \Pi_{c_3c'_3c_4c'_4}(X_3,X'_3;X_4,X'_4) \calY_{c_4c'_4c_2c'_2}(X_4,X'_4;X_2,X'_2) .
\end{multline}
Using~(\ref{phi2nu}), we obtain 
\begin{multline} 
\calY^{\nu_1\nu_2}_{c_1c'_1c_2c'_2}(x_1,x'_1;x_2,x'_2) = \calX^{\nu_1\nu_2}_{c_1c'_1c_2c'_2}(x_1,x'_1;x_2,x'_2)
 \\ - \quarter \sum_{c_3\cdots c'_4 \atop \nu_3,\nu_4} \int dx_3 dx'_3 dx_4 dx'_4 \, \calX^{\nu_1\nu_3}_{c_1c'_1c_3c'_3}(x_1,x'_1;x_3,x'_3)  \\ \times \Pi^{\nu_3\nu_4}_{c_3c'_3c_4c'_4}(x_3,x'_3;x_4,x'_4) \calY^{\nu_4\nu_2}_{c_4c'_4c_2c'_2}(x_4,x'_4;x_2,x'_2) .
\end{multline}
In Fourier space, introducing the total and relative momentum-frequency of the pair [Eqs.~(\ref{appbs1},\ref{appbs2})], we then obtain~(\ref{eqbs1}). Equation~(\ref{eqbs1a}) is derived in a similar way.

\section{Initial condition of the RG flow in the Hubbard model}
\label{app_init}

In this Appendix, we discuss in more detail the initial condition of the RG flow for the two-dimensional  Hubbard model at half-filling. From Eq.~(\ref{gap1b}), we deduce 
\begin{equation}
\bar\Sig_{\rm sp}^z(p,p') = - \delta_{p',p+Q} m 
\end{equation}
and 
\begin{equation}
\begin{split}
\bGch(p,p') &= - \delta_{p,p'} \frac{2(i\wn+\eps_\p)}{\wn^2+\Ep^2} \equiv \delta_{p,p'} \bGch(p) , \\ 
\bGsp(p,p') &=  \delta_{p+Q,p'} \frac{2m}{\wn^2+\Ep^2}  \equiv \delta_{p+Q,p'} \bGsp(p),
\end{split}
\end{equation}
where 
\begin{equation}
\eps_\p = -2t (\cos p_x + \cos p_y) 
\label{epsp}
\end{equation}
($t$ denotes the nearest-neighbor hopping amplitude), $\Ep=(\eps_\p^2+m^2)^{1/2}$,  $Q=(\Q,0)$ and $\Q=(\pi,\pi)$. The gap equation~(\ref{gap1a}) then leads to~(\ref{gap2}). 

Since the 2PI vertex is momentum and frequency independent for $k=\Lamb$, the Bethe-Salpeter equation for the pair propagator $\bar W^{(2)}\equiv\bar\Gamma^{(2)-1}$ is exactly solvable. Focusing on the transverse spin channel, we obtain 
\begin{equation}
\begin{split}
&\bar W_{\rm sp}^{(2)xx}(q) = \frac{1}{\bar D(q)} \bigl\lbrace \bar\Pi^{xx}_{\rm sp}(q) [ 1 + \Usp \bar\Pi^{xx}_{\rm sp}(q+Q) ] \\ & \hspace{2cm} + \Usp \bar\Pi^{xy}_{\rm sp}(q,q+Q)^2 \bigr\rbrace , \\ 
&\bar W_{\rm sp}^{(2)xy}(q,q+Q) = \frac{1}{\bar D(q)}\bar\Pi^{xy}_{\rm sp}(q,q+Q) ,
\end{split}
\label{appinit1}
\end{equation} 
where 
\begin{align}
\bar D(q) ={}& [ 1 + \Usp \bar \Pi^{xx}_{\rm sp}(q) ] [ 1 + \Usp \bar\Pi^{xx}_{\rm sp}(q+Q) ]  \nonumber \\ & + \Usp^2 \bar\Pi^{xy}_{\rm sp}(q,q+Q)^2
\label{appinit2}
\end{align}
and
\begin{equation}
\begin{gathered}
\bar\Pi_{\rm sp}^{xx}(q) = - \half \int_l [ \bGch(l_-) \bGch(l_+) - \bGsp(l_-) \bGsp(l_+) ] , \\ 
\bar\Pi_{\rm sp}^{xy}(q,q+Q) = \frac{i}{2} \int_l [ \bGch(l_-) \bGsp(l_+) - \bGsp(l_-) \bGch(l_+) ]
\end{gathered}  
\label{Pixx} 
\end{equation}
We use the notation $\int_l=T\sum_{\wn} \int_\l$ and $l_\pm=l\pm q/2$. The dispersion of the collective spin modes is obtained from the poles of $\bar W^{(2)}$, i.e. from the zeros of $\bar D(q)$,  after the analytical continuation $q=(\q,i\wnu)\to (\q,\w+i0^+)$. Using 
\begin{equation} 
1+\Usp \bar\Pi^{xx}_{\rm sp}(Q) = 0 
\label{goldinit} 
\end{equation}
(which is a consequence of the gap equation~(\ref{gap2})), we find 
\begin{equation}
\bar D(q) = 4[A(\q-\Q)^2 + B \wnu^2] 
\end{equation}
to order $(\q-\Q)^2$ and $\wnu^2$, with 
\begin{equation}
\begin{split} 
A ={}& \int_\p \eps^2 \calN_2 \int_\p [\calN_2(2\eps_1^2+\eps\eps_2)- 4 \eps^2\eps_1^2 \calN_3 ] , \\ 
B ={}& -\int_\p \eps^2 \calN_2 \int_\p (2 \calN_2 - 4 E^2 \calN_3) \\ & + 16 m^2 \left( \int_\p (\calN_2 - E^2 \calN_3) \right)^2  ,
\end{split}
\end{equation}
where $\eps=\eps_\p$, $\eps_1=\partial_{p_x}\eps$, $\eps_2=\partial^2_{p_x}\eps$, $E=\Ep$, and  
\begin{equation}
\calN_i = \frac{1}{\beta} \sum_{\wn} \frac{1}{(\wn^2+E)^i} . 
\end{equation}
For $T=0$, we use $\calN_2=1/4E^3$ and $\calN_3=3/16E^5$ to obtain 
\begin{equation}
\begin{split}
A &= \llangle\frac{\eps^2}{4E^3}\rrangle \llangle \frac{2\eps_1^2+\eps\eps_2}{4E^3}  - \frac{3\eps^2 \eps_1^2}{4E^5} \rrangle , \\ 
B &= \frac{1}{16} \llangle \frac{1}{E^3} \rrangle \left( \llangle\frac{\eps^2}{E^3}\rrangle  + m^2 \llangle \frac{1}{E^3} \rrangle \right) ,
\end{split}
\end{equation}
where we use the notation $\langle\cdots\rangle=\int_\p\cdots$. With an integration by part, we find
\begin{equation}
3 \llangle \frac{\eps^2\eps_1^2}{E^5} \rrangle = \llangle \frac{\eps_1^2+\eps\eps_2}{E^3} \rrangle 
\end{equation}
and 
\begin{equation}
A = \frac{1}{16} \llangle\frac{\eps^2}{E^3}\rrangle  \llangle\frac{\eps_1^2}{E^3}\rrangle .
\end{equation}
The velocity $c=\sqrt{A/B}$ agrees with the known RPA result [Eq.~(\ref{velocity})]. In the large $U$ limit, using $m\simeq U/2$, $\langle\eps^2\rangle=4t^2$ and $\langle\eps_1^2\rangle=2t^2$, we finally obtain $c=\sqrt{2}J$. 

\section{Ward identity and Goldstone's theorem}
\label{app_goldstone}

Let us consider an infinitesimal transformation,
\begin{equation}
\psi_\alpha \to \psi_\alpha + \eps \sum_{\alpha'} \calT_{\alpha\alpha'} \psi_{\alpha'} + \calO(\eps^2),
\label{agold1} 
\end{equation}
which leaves the action $S^{(0)}+\Sint+\Delta S_k$ invariant. The invariance of the partition function $Z_k[J]$ in the change of variable~(\ref{agold1}) then implies  
\begin{equation}
\int \calD[\psi] \, e^{-S^{(0)}-\Sint-\Delta S_k} \delta S_J = 0  
\end{equation}
to order $\eps$, where 
\begin{equation}
\delta S_J = - \eps \sum_{\alpha,\alpha',\alpha''} \calT_{\alpha'\alpha''} J_{\alpha\alpha'} \psi_\alpha \psi_{\alpha''} 
\end{equation}
gives the variation of the source term $S_J=-\half  \sum_{\alpha,\alpha'} \psi_\alpha J_{\alpha\alpha'} \psi_{\alpha'}$. Here we assume that the transformation~(\ref{agold1}) is free of anomaly, i.e. that its Jacobian is a constant. We thus obtain 
\begin{equation}
\sum_{\alpha,\alpha',\alpha''} \calT_{\alpha'\alpha''} J_{\alpha\alpha'} W^{(1)}_{k,\alpha\alpha''} = 0 . 
\end{equation}
The functional derivative wrt $J_{\gam_2}$ yields 
\begin{align}
& \sum_{\gam_1,\alpha''_1} \calT_{\alpha'_1\alpha''_1} J_{\gam_1} W^{(2)}_{k;\alpha_1\alpha''_1,\gam_2}
\nonumber \\ + & \sum_{\alpha''_1} \bigl( \calT_{\alpha'_2\alpha''_1} W^{(1)}_{k,\alpha_2\alpha''_1} - \calT_{\alpha_2\alpha''_1} W^{(1)}_{k,\alpha'_2\alpha''_1} \bigr) = 0 .
\end{align}
In the equilibrium state, the source takes the value $J_\gam= - \half \sum_{\gam'} R_{k,\gam\gam'} \bar\calG_{k,\gam'}$ (Sec.~\ref{subsec_eqstate}), which gives the following Ward identity 
\begin{align}
& \half \sum_{\gam_1,\gam_3,\alpha''_1} \calT_{\alpha'_1\alpha''_1} R_{k,\gam_1\gam_3} \bar W^{(2)}_{k;\alpha_1\alpha''_1,\gam_2} \bar\calG_{k,\gam_3} \nonumber \\ 
+ & \sum_{\alpha''_1} \bigl( \calT_{\alpha'_2\alpha''_1} \bar\calG_{k,\alpha_2\alpha''_1} - \calT_{\alpha_2\alpha''_1} \bar\calG_{k,\alpha'_2\alpha''_1} \bigr) = 0 .
\label{agold2} 
\end{align}

Let us now consider a system with antiferromagnetic long-range order polarized along the $z$ axis. We assume that the order takes place in the $s$-wave ph channel and choose a cutoff function of the form
\begin{align}
R^{\nu_1\nu_2}_{k,\rm sp}(x_1,x'_1;x_2,x'_2) ={}& \delta_{\nu_1,\nu_2}\delta(x_1-x'_1{}^+) \delta(x_2-x'_2{}^+) \nonumber \\ & \times R_{k,\rm sp}(x_1-x_2).
\label{agold4} 
\end{align} 
An infinitesimal spin rotation about the $x$ axis corresponds to 
\begin{equation}
\calT_{cc'}(X,X') = \frac{i}{2} \delta_{c,c'} \delta(x-x') c \sig^{x}_{\sig\sig'} .
\end{equation}
With $x_2'=x_2^-$, $c_2=\bar c_2'=+$ and $\sig_2=\bar\sig'_2=\up$ (we use the notation $\bar c=-c$ and $\bar\sig=-\sig$, i.e. $\bar\sig=\down$ if $\sig=\up$ and {\it vice versa}), Eq.~(\ref{agold2}) gives 
\begin{multline}
0 = \half \sum_{c_1,c'_1,c_3,c'_3} \sum_{\sig_1,\sig'_1,\sig_3,\sig'_3 \atop \sig''_1} \int dx_1 dx'_1 dx_3 dx'_3 \, c'_1 \sig^x_{\sig'_1\sig''_1} \\
\times R_{k,c_1c'_1c_3c'_3}(X_1,X'_1;X_3,X'_3) \bar\calG_{k,c_3c'_3}(X_3,X'_3) \\
\times \bar W^{(2)}_{k;c_1\sig_1,c'_1\sig''_1,+\up,-\down}(x_1,x'_1;x_2^+,x_2) \\
- \sum_{\sig''_1} [ \sig^x_{\down\sig''_1} \bar\calG_{k;+\up,-\sig''_1}(x^+_2,x_2) + \sig^x_{\up\sig''_1} \bar\calG_{k;-\down,+\sig''_1}(x_2,x^+_2) ] . 
\label{agold3} 
\end{multline}
Performing the sum over $\sig''_1$ in the last term, we obtain 
\begin{align} 
& - \bar\calG_{k;+\up,-\up}(x^+_2,x_2) - \bar\calG_{k;-\down,+\down}(x_2,x^+_2) \nonumber \\ 
={}& \bar\calG_{k;-\up,+\up}(x_2,x^+_2) - \bar\calG_{k;-\down,+\down}(x_2,x^+_2) \nonumber \\
={}& \bar\calG_{k,-+}^z(x_2,x^+_2)  . 
\label{agold3a} 
\end{align} 
As for the first term in~(\ref{agold3}), we note that we must have $c_1=\bar c'_1$ and $c_3=\bar c'_3$, $\sig''_1=\bar\sig'_1$ and $\sig_1=\bar\sig''_1$, which in turn implies $\sig_3=\sig'_3$. Using~(\ref{agold4}), we can then rewrite this term as 
\begin{align}
& \sum_{c_1,\sig_1,\sig_3} \int dx_1 dx_3 \, \bar c_1
R_{k;c_1\sig_1,\bar c_1\sig_1,+\sig_3,-\sig_3}(x_1-x_3) \nonumber \\ & \times \bar\calG_{k;+\sig_3,-\sig_3}(x^+_3,x_3) \bar W^{(2)}_{k;c_1\sig_1,\bar c_1\bar \sig_1,+\up,-\down}(x_1,x_2) ,
\end{align}
where the spin correlation function $\bar W^{(2)}_{k;c_1\sig_1,c'_1\sig'_1,c_2\sig_2,c'_2\sig'_2}(x_1,x_2)$ is defined by~(\ref{Wsp}). Since 
\begin{align}
& \sum_{c_1} \bar c_1 R_{k;c_1\sig_1,\bar c_1\sig_1;+\sig_3,-\sig_3}\bar W^{(2)}_{k;c_1\sig_1,\bar c_1\bar \sig_1,+\up,-\down} \nonumber \\ ={}&
- 2 \sig_1 \sig_3 R_{k,\rm sp} \bar W^{(2)}_{k;+\sig_1,-\bar\sig_1,+\up,-\down} ,
\end{align}
and 
\begin{align}
& \sum_{\sig_1} \sig_1 \bar W^{(2)}_{k;+\sig_1,-\bar\sig_1,+\up,-\down}(x_1,x_2) \nonumber \\ 
={}& -\half [ \bar W^{(2)yy}_{k,\rm sp}(x_1,x_2) - i  \bar W^{(2)yx}_{k,\rm sp}(x_1,x_2) ]
\end{align}
we finally obtain 
\begin{align} 
& - \int dx_1 dx_3 R_{k,\rm sp}(x_1-x_3) \bar\calG^z_{k,-+}(x_3,x^+_3) \nonumber \\ & 
\times [ \bar W^{(2)yy}_{k,\rm sp}(x_1,x_2) - i  \bar W^{(2)yx}_{k,\rm sp}(x_1,x_2) ] 
\label{agold5}
\end{align}
for the first term of~(\ref{agold3}). From~(\ref{agold3a}) and (\ref{agold5}), we deduce 
\begin{align} 
& \bar\calG^z_{k,-+}(x_2,x^+_2) - \int dx_1 dx_3 R_{k,\rm sp}(x_1-x_3) \bar\calG^z_{k,-+}(x_3,x^+_3) \nonumber \\ & 
\times [ \bar W^{(2)yy}_{k,\rm sp}(x_1,x_2) - i  \bar W^{(2)yx}_{k,\rm sp}(x_1,x_2) ] = 0. 
\end{align}
Since $\bar\calG^z_{k,-+}(x,x^+)=M(-1)^\r$ in the linearly polarized antiferromagnetic state, we find 
\begin{equation}
1 - R_{k,\rm sp}(Q) \bar W^{(2)yy}_{k,\rm sp}(Q,Q) = 0 ,
\label{agold6}
\end{equation} 
where we have used $\bar W^{(2)yy}_{k,\rm sp}(q,q')\propto \delta_{q,q'}$, $\bar W^{(2)yx}_{k,\rm sp}(q,q')\propto \delta_{q+Q,q'}$, and $\bar W^{(2)yx}_{k,\rm sp}(Q,0)=0$. 

If we assume that the 2PI vertex $\bar\Phi^{(2)xx}_{k,\rm sp}=u_{k,\rm sp}$ in the transverse spin channel is momentum and frequency independent (apart from conservation of momentum and frequency) then the Bethe-Salpeter equation satisfied by $\bar W^{(2)}_k$ can be easily solved,
\begin{align}
& \bar W_{k,\rm sp}^{(2)xx}(q,q) = \frac{1}{\bar D_k(q)} \bigl\lbrace \bar \Pi_{k,\rm sp}^{xx}(q) [ 1 + \bar \calX_{k,\rm sp}^{xx}(q+Q) \nonumber \\ & \times \bar \Pi_{k,\rm sp}^{xx}(q+Q) ]  + \bar \calX_{k,\rm sp}^{xx}(q+Q) \bar \Pi_{k,\rm sp}^{xy}(q,q+Q)^2 \bigr\rbrace ,  \nonumber \\ 
& \bar W_{k,\rm sp}^{(2)xy}(q,q+Q) = \frac{1}{\bar D_k(q)}\bar \Pi_{k,\rm sp}^{xy}(q,q+Q) ,
\end{align}
where 
\begin{align}
\bar D_k(q) ={}& [ 1 + \bar \calX_{k,\rm sp}^{xx}(q) \bar \Pi_{k,\rm sp}^{xx}(q) ] \nonumber \\ & \times [ 1 + \bar \calX_{k,\rm sp}^{xx}(q+Q) \bar \Pi_{k,\rm sp}^{xx}(q+Q) ]   \\ & + \bar \calX_{k,\rm sp}^{xx}(q)\bar \calX_{k,\rm sp}^{xx}(q+Q)\bar \Pi_{k,\rm sp}^{xy}(q,q+Q)^2 , \nonumber
\end{align}
and $\bar\calX^{xx}_{k,\rm sp}(q)=R_{k,\rm sp}(q)+u_{k,\rm sp}$.
$\bar \Pi_{k,\rm sp}^{xx}(q,q)$ and $\bar \Pi_{k,\rm sp}^{xy}(q,q+Q)$ are defined as in Eqs.~(\ref{Pixx}). Since $\bar\Pi_k^{xy}(0,Q)=0$ and 
\begin{equation}
\bar W_{k,\rm sp}^{(2)xx}(Q,Q) = \frac{\bar  \Pi_{k,\rm sp}^{xx}(Q) }{ 1 + \bar \calX_{k,\rm sp}^{xx}(Q) \bar \Pi_{k,\rm sp}^{xx}(Q) } ,
\label{Wgold}
\end{equation}
the Ward identity~(\ref{agold6}) implies 
\begin{equation}
1 + u_{k,\rm sp} \bar \Pi_{k,\rm sp}^{xx}(Q) = 0 ,
\end{equation}
which agrees with~(\ref{goldinit}) for $k=\Lamb$.

\section{RG equations for the 2PI vertices} 
\label{app_rgeq} 

In this Appendix, we derive the RG equations~(\ref{dphi1}) and (\ref{dphi2}). 

\subsection{One-particle vertex: $\dk\Phi_k^{(1)}[\calG]$} 

Using Eq.~(\ref{rgeq1}) and 
\begin{equation}
\Gamma^{(3)}_{k,\gam_1\gam_2\gam_3} = \Phi^{(3)}_{k,\gam_1\gam_2\gam_3} + \frac{\delta \Pi^{-1}_{\gam_2\gam_3}}{\delta \calG_{\gam_1}} ,
\label{apprgeq1a}
\end{equation}
we obtain 
\begin{align}
\dk \Phi_{k,\gam_1}^{(1)} ={}& \frac{1}{3!} \tdk \Tr\biggl\lbrace W_k^{(2)} \biggl[ \Phi_{k,\gam_1}^{(3)} + \frac{\delta \Pi^{-1}}{\delta \calG_{\gam_1}} \biggr] \biggr\rbrace \nonumber \\ & - \third \sum_{\gam_2} \dot R_{k,\gam_1\gam_2} \calG_{\gam_2} .
\label{apprgeq1}
\end{align} 
We then use 
\begin{align}
\Tr\biggl[ W_k^{(2)} \frac{\delta \Pi^{-1}}{\delta \calG_{\gam_1}} \biggr]  
&= - \Tr\biggl[ \Pi(\calI+\calX_k\Pi)^{-1} \Pi^{-1} \frac{\delta\Pi}{\delta\calG_{\gam_1}} \Pi^{-1}\biggr]  \nonumber \\ 
&= -  \Tr\biggl[ (\Pi^{-1} - \calY_k )  \frac{\delta\Pi}{\delta\calG_{\gam_1}} \biggr] .
\end{align}
With
\begin{align}
\frac{\delta\Pi_{\gam_2\gam_3}}{\delta\calG_{\gam_1}} ={}& - \calI_{\gam_1,\alpha_2\alpha_3} \calG_{\alpha'_2\alpha'_3} - \calI_{\gam_1,\alpha'_2\alpha'_3} \calG_{\alpha_2\alpha_3} \nonumber \\ & 
+ \calI_{\gam_1,\alpha_2\alpha'_3} \calG_{\alpha'_2\alpha_3} + \calI_{\gam_1,\alpha'_2\alpha_3} \calG_{\alpha_2\alpha'_3} ,
\label{apprgeq1b}
\end{align}
this gives 
\begin{align}
\Tr\biggl[ W_k^{(2)} \frac{\delta \Pi^{-1}}{\delta \calG_{\gam_1}} \biggr] 
={}& - \Tr\biggl[ \Pi^{-1} \frac{\delta\Pi}{\delta\calG_{\gam_1}} \biggr] \nonumber \\ &
+ 2 \sum_{\gam_2} \calG_{\gam_2} \calY_{k;\alpha_1\alpha_2,\alpha'_2\alpha'_1} .
\label{apprgeq2}
\end{align} 
Since the first term in the rhs of~(\ref{apprgeq2}) does not depend on $k$, Eq.~(\ref{apprgeq1}) yields Eq.~(\ref{dphi1}) (we use $\tdk\calX_k=\dot R_k$).

\subsection{Two-particle vertex: $\dk\Phi_k^{(2)}[\calG]$} 

From Eq.~(\ref{apprgeq1a}) and 
\begin{equation}
\Gamma^{(4)}_{k,\gam_1\gam_2\gam_3\gam_4} = \Phi^{(4)}_{k,\gam_1\gam_2\gam_3\gam_4} + \frac{\delta^2 \Pi^{-1}_{\gam_3\gam_4}}{\delta \calG_{\gam_1}\delta \calG_{\gam_2}} ,
\end{equation} 
we obtain 
\begin{align}
\dk \Phi_{k,\gam_1\gam_2}^{(2)} ={}& \frac{1}{3!} \tdk \Tr\biggl[ \frac{\delta W_k^{(2)}}{\delta\calG_{\gam_2}} \Phi_{k,\gam_1}^{(3)} + W_k^{(2)} \Phi_{k,\gam_1\gam_2}^{(4)} \biggr] \nonumber \\ &
+ \third \tdk [ \Delta\calY_{k;\alpha_1\alpha_2,\alpha'_2\alpha'_1} - (\alpha_2 \leftrightarrow \alpha'_2) ] \nonumber \\ &
+ \third \tdk \sum_{\gam_3} \calG_{\gam_3} \frac{\delta}{\delta\calG_{\gam_2}} \Delta\calY_{k;\alpha_1\alpha_3,\alpha'_3\alpha'_1} .
\label{apprgeq3}
\end{align} 

Using $W_k^{(2)}=(\Pi^{-1}+\calX_k)^{-1}$ and 
\begin{align}
\frac{\delta W_k^{(2)}}{\delta\calG_{\gam}} &= - W_k^{(2)} \biggl( \Phi^{(3)}_{k,\gam} - \Pi^{-1} \frac{\delta \Pi}{\delta \calG_\gam} \Pi^{-1} \biggr) W_k^{(2)} \nonumber \\ 
&=  - W_k^{(2)} \Phi^{(3)}_{k,\gam} W_k^{(2)} + \calJ_k \frac{\delta \Pi}{\delta \calG_\gam} \calJ_k^T
\end{align}
($\calJ$ and $\calJ^T$ are defined in~(\ref{Jdef})) and~(\ref{apprgeq1b}), we obtain 
\begin{multline}
\frac{\delta W_{k,\gam_3\gam_4}^{(2)}}{\delta\calG_{\gam_2}} =  - \bigl( W_k^{(2)} \Phi^{(3)}_{k,\gam_2}W_k^{(2)} \bigr)_{\gam_3\gam_4} \\
+ \sum_{\gam_5} \calG_{\gam_5} \bigl[ \calJ_{k;\gam_3,\alpha_2\alpha_5} \calJ^T_{k;\alpha'_5\alpha'_2,\gam_4} - (\alpha_2 \leftrightarrow \alpha'_2) \bigr] . 
\label{apprgeq4}
\end{multline} 
Furthermore, Eq.~(\ref{apprgeq4}) with $\Delta\calY_k=-\calX_k W_k^{(2)} \calX_k$ and $\calX_k \calJ_k = \calJ_k^T \calX_k = \calY_k$ implies
\begin{multline}
\frac{\delta}{\delta\calG_{\gam_2}} \Delta\calY_{k;\alpha_1\alpha_3,\alpha'_3\alpha'_1} = \\ 
- \bigl( \Phi^{(3)}_{k,\gam_2} W_k^{(2)} \calX_k + \calX_k  W_k^{(2)} \Phi^{(3)}_{k,\gam_2} \bigr)_{\alpha_1\alpha_3,\alpha'_3\alpha'_1} \\ 
+ \quarter \sum_{\gam_4,\gam_5} \calX_{k;\alpha_1\alpha_3,\gam_4} \bigl( W_k^{(2)} \Phi^{(3)}_k W_k^{(2)} \bigr)_{\gam_4\gam_5} \calX_{k;\gam_5,\alpha'_3\alpha'_1} \\
- \sum_{\gam_6} \calG_{\gam_6} [ \calY_{k;\alpha_1\alpha_3,\alpha_2\alpha_6} \calY_{k;\alpha'_6\alpha'_2,\alpha'_3\alpha'_1} - (\alpha_2 \leftrightarrow \alpha'_2) ] . 
\end{multline} 
Using finally $W_k^{(2)}\calX_k-\calI=-\calJ_k$ and $\calX_k W_k^{(2)}-\calI=-\calJ^T_k$, we conclude that 
\begin{multline}
\frac{\delta}{\delta\calG_{\gam_2}} \Delta\calY_{k;\alpha_1\alpha_3,\alpha'_3\alpha'_1} = 
- \bigl( \Phi^{(3)}_{k,\gam_2} - \calJ_k^T \Phi^{(3)}_{k,\gam_2} \calJ_k \bigr)_{\alpha_1\alpha_3,\alpha'_3\alpha'_1} \\ 
- \sum_{\gam_4} \calG_{\gam_4} [ \calY_{k;\alpha_1\alpha_3,\alpha_2\alpha_4} \calY_{k;\alpha'_4\alpha'_2,\alpha'_3\alpha'_1} - (\alpha_2 \leftrightarrow \alpha'_2) ] . 
\label{apprgeq5}
\end{multline}
From Eqs.~(\ref{apprgeq3}), (\ref{apprgeq4}) and (\ref{apprgeq5}) we deduce~(\ref{dphi2}) (using $\tdk \Phi^{(3)}_k=0$).

\section{Luttinger-Ward functional and symmetries}
\label{app_phi_sym} 

In this Appendix we show how the symmetries of the action constrain the perturbative expansion of the Luttinger-Ward functional $\Phi[\calG]=\Phi_{k=0}[\calG]$ in the Hubbard model. To order $U$, $\Phi[\calG]$ is given by~(\ref{Phi_2}), i.e.  
\begin{align}
&\frac{1}{32} \sum_{c_1\cdots c'_2 \atop \nu_1,\nu_2,\mu} \int dx\, U^{\mu\mu}_{c_1c'_1c_2c'_2} \calG^{\nu_1}_{c_1c'_1}(x,x) \calG^{\nu_2}_{c_2c'_2}(x,x) \nonumber \\ & \times  \tr\bigl(\tau^{\nu_1}_{c_1c'_1}\tau^{\mu\dagger}_{c_1c'_1}\bigr) 
\tr\bigl(\tau^{\nu_2}_{c_2c'_2}\tau^{\mu\dagger}_{c_2c'_2}\bigr) ,
\label{appphi1}
\end{align}
where $U^{\mu\mu}_{c_1c'_1,c_2c'_2}$ is defined by~(\ref{Udef1},\ref{Udef2}) and the symmetry properties~(\ref{fullysym}). Performing the traces in~(\ref{appphi1}), we find 
\begin{equation}
\frac{U}{4} \int dx \bigl[ \Fs^\dagger(x,x) \Fs(x,x) + \Gch(x,x^+)^2 - \Gspv(x,x^+)^2 \bigr] . 
\label{appphi2}
\end{equation}
As expected, it is possible to express the result in terms of the invariants~(\ref{invdef}). One of the invariants does not appear since the triplet component $U_{\rm t}$ of the interaction vanishes in the Hubbard model. 

Similarly, the $\calO(U^2)$ contribution to the Luttinger-Ward functional can be written as
\begin{align}
&- \frac{1}{48\times 2^4} \sum_{c_1\cdots c'_4} \sum_{\nu_1\cdots \nu_4 \atop \mu,\mu'} \int dx dy\, U^{\mu\mu}_{c_1c_2c_3c_4} U^{\mu'\mu'}_{c'_1c'_2c'_3c'_4} \nonumber \\ & \times  \calG^{\nu_1}_{c_1c'_1}(x,y) \calG^{\nu_2}_{c_2c'_2}(x,y) \calG^{\nu_3}_{c_3c'_3}(x,y) \calG^{\nu_4}_{c_4c'_4}(x,y)  \nonumber \\ & \times 
\tr \bigl( \tau^{\mu*}_{c_1c_2} \tau^{\nu_2}_{c_2c'_2} \tau^{\mu'\dagger}_{c'_1c'_2} \tau^{\nu_1T}_{c_1c'_1} \bigr) 
\tr \bigl( \tau^{\mu*}_{c_3c_4} \tau^{\nu_4}_{c_4c'_4} \tau^{\mu'\dagger}_{c'_3c'_4} \tau^{\nu_3T}_{c_3c'_3} \bigr) ,
\end{align} 
which eventually gives 
\begin{widetext}
\begin{align} 
& - \frac{U^2}{32} \int dxdy \bigl\lbrace \bigl[ \Gspv(x,y)^2-\Gch(x,y)^2 \bigr] \bigl[ \Gspv(y,x)^2-\Gch(y,x)^2 \bigr] + \bigl[ \Ftv^\dagger(x,y)^2-\Fs^\dagger(x,y)^2 \bigr] \bigl[ \Ftv(x,y)^2-\Fs(x,y)^2 \bigr] \bigr\rbrace \nonumber \\ &
+ \frac{U^2}{16} \int dxdy \bigl\lbrace 
- \bigl[ \Ftv^\dagger(x,y) \cdot \Ftv(x,y) \bigr] \bigl[ \Gspv(x,y) \cdot \Gspv(y,x) \bigr]
+ 2 \bigl[\Ftv^\dagger(x,y) \cdot \Gspv(x,y) \bigr] \bigl[\Ftv(x,y) \cdot \Gspv(y,x) \bigr] \nonumber \\ &
+ \bigl[ \Gspv(x,y) \cdot \Gspv(y,x) \bigr] \Fs^\dagger(x,y) \Fs(x,y)
- 2 \bigl[ \Ftv^\dagger(x,y) \cdot \Gspv(y,x) \bigr] \Fs(x,y) \Gch(x,y) \\ &
- 2 \bigl[ \Ftv(x,y) \cdot \Gspv(y,x) \bigr] \Fs^\dagger(x,y) \Gch(x,y) 
- \bigl[ \Ftv^\dagger(x,y) \cdot \Ftv(x,y) + \Fs^\dagger(x,y) \Fs(x,y) \bigr] \Gch(x,y)\Gch(y,x) \nonumber \\ &
+ 2i \Ftv(x,y) \cdot \bigl[ \Ftv^\dagger(x,y) \times \Gspv(y,x) \bigr]  \Gch(x,y)
+ i  \bigl[ \Fs(x,y)\Ftv^\dagger(x,y) -  \Fs^\dagger(x,y) \Ftv(x,y) \bigr] \cdot \bigl[ \Gspv(x,y) \times \Gspv(y,x) \bigr] \nonumber 
\bigr\rbrace .
\end{align}
\end{widetext}
All terms are obviously SU(2) spin-rotation and U(1) invariant. Some of them can be expressed as a function of the quadratic invariants~(\ref{invdef}), but there are also quartic (i.e. $\calO(\calG^4)$) invariants. 

It should be noticed that the Luttinger-Ward functional $\Phi[\calG]\equiv \Phi_{k=0}[\calG]$ has a higher degree of symmetry than the scale-dependent Luttinger-Ward functional $\Phi_k[\calG]$. This comes from the fact that in the Hubbard model, the interaction action $\Sint$ is invariant in a local SU(2) spin rotation, 
\begin{equation}
\left(
\begin{array}{cc}
\psi_{c\up}(x) \\
\psi_{c\down}(x) 
\end{array}
\right) 
\to e^{\frac{i}{2} c \eps(x)\sigbf \cdot \n}
\left(
\begin{array}{cc}
\psi_{c\up}(x) \\
\psi_{c\down}(x) 
\end{array}
\right) 
\label{appsym1}
\end{equation}
(with $\n$ an arbitrary unit vector and $\eps(x)$ an arbitrary function of $x$), and a local U(1) transformation 
\begin{equation}
\psi_{c\sig}(x) \to e^{- i c \eps(x)} \psi_{c\sig}(x) .
\label{appsym2}
\end{equation}
This implies that $\Phi[\calG]$ is invariant in the local transformations 
\begin{align}
\calG_{c\sig,c'\sig'}(x,x') \to{}& \sum_{\sig_1,\sig'_1} \bigl( e^{ \frac{i}{2} c \eps(x)\sigbf \cdot \n} \bigr)_{\sig\sig_1} \calG_{c\sig_1,c'\sig'_1}(x,x') \nonumber \\ & \times \bigl( e^{ \frac{i}{2} c' \eps(x')\sigbf \cdot \n} \bigr)_{\sig'\sig'_1}  
\label{appsym3}
\end{align}
and
\begin{equation}
\calG_{c\sig,c'\sig'}(x,x') \to e^{- i c \eps(x)} \calG_{c\sig,c'\sig'}(x,x') e^{- i c' \eps(x')} .
\label{appsym4}
\end{equation}
In general, however, the regulator term $\Delta S_k$ is not invariant in the transformations~(\ref{appsym1}) and (\ref{appsym2}) if $\eps(x)$ is time or space dependent, so that $\Phi_k[\calG]$ ($k>0$) is invariant in the transformations~(\ref{appsym3}) and (\ref{appsym4}) only for $\eps(x)=\const$. Moreover, once we expand $\Phi_k[\calG]$ about a nontrivial set of minima ($\bar\calG_k\neq 0$), we loose the invariance under the local transformations~(\ref{appsym3}) and (\ref{appsym4}). Even when the regulator term $\Delta S_k$ is invariant in the transformations~(\ref{appsym1}) and (\ref{appsym2}), the expanded functional $\Phi_k[\calG]$ remains invariant in the local transformations~(\ref{appsym3}) and (\ref{appsym4}) only if we simultaneously transform the 2PI vertices. For instance, with the quadratic ansatz~(\ref{Phi_5}) discussed in Sec.~\ref{subsec_approxsol}, one should transform the two-particle 2PI vertex as 
\begin{equation}
\begin{split}
u_{k,\rm ph}^{\nu\nu}(x_1,x'_1;x_2,x'_2) \to{}& u_{k,\rm ph}^{\nu\nu}(x_1,x'_1;x_2,x'_2) \\ & \times  e^{i[\eps(x_1)-\eps(x'_1)+\eps(x_2)-\eps(x'_2)]} , \\
u_{k,\rm pp}^{\nu\nu}(x_1,x'_1;x_2,x'_2) \to{}& u_{k,\rm pp}^{\nu\nu}(x_1,x'_1;x_2,x'_2) \\ & \times  e^{i[\eps(x_1)+\eps(x'_1)-\eps(x_2)-\eps(x'_2)]}, 
\end{split}
\end{equation}
in the gauge transformation~(\ref{appsym2}).

%



\begin{thebibliography}{79}
\bibitem [{not({\natexlab{a}})}]{note11}%
  \BibitemOpen
  \href@noop {} {} \bibinfo {note} {One-dimensional systems
  are special in this respect since the Fermi surface reduces to two Fermi
  points $\pm p_F$.}\BibitemShut {Stop}%
\bibitem [{not({\natexlab{c}})}]{note15}%
  \BibitemOpen
  \href@noop {} {} \bibinfo {note} {It has been proposed to
  suppress divergences associated to a phase transition by introducing a small
  symmetry-breaking field, see
  Refs.~\onlinecite{Gersch05,Salmhofer04,Maier12}.}\BibitemShut {Stop}%
\bibitem [{not({\natexlab{d}})}]{note1}%
  \BibitemOpen
  \href@noop {} {} \bibinfo {note} {Since there is in
  general no ambiguity, we use the symbol $U$ both for the function
  $U_{\alpha_1\alpha_2\alpha_3\alpha_4}$ [Eq.~(\ref{action0})] and for the
  onsite interaction in the Hubbard model.}\BibitemShut {Stop}%
\bibitem [{not({\natexlab{e}})}]{note6}%
  \BibitemOpen
  \href@noop {} {} \bibinfo {note} {If the band energy
  satisfies $|\xi_\p|\leq\eps_{k=\Lamb}$, then the model with action $S+\Delta
  S_\Lamb$ and hard cutoff~(\ref{G0k}) is exactly (trivially) solvable
  regardless of the cutoff function $R_k$.}\BibitemShut {Stop}%
\bibitem [{not({\natexlab{f}})}]{note2}%
  \BibitemOpen
  \href@noop {} {} \bibinfo {note} {The fermionic trace is
  defined by $\tr\,A=\sum_\alpha A_{\alpha\alpha}$ for any ``fermionic" matrix
  $A$, e.g. $\tr\,\calG^{(0)}=\sum_\alpha
  \calG^{(0)}_{\alpha\alpha}$.}\BibitemShut {Stop}%
\bibitem [{not({\natexlab{g}})}]{note12}%
  \BibitemOpen
  \href@noop {} {} \bibinfo {note} {This property follows
  from the fact that $\calG_k^{(0)-1}$ can be absorbed in the source term in
  Eq.~(\ref{Zdef}).}\BibitemShut {Stop}%
\bibitem [{not({\natexlab{h}})}]{note7}%
  \BibitemOpen
  \href@noop {} {} \bibinfo {note} {Note that the triplet
  vector field $\O_{\rm
  t}(x,x')=(O^x_{--}(x,x'),O^y_{--}(x,x'),O^z_{--}(x,x'))^T$ transforms as a
  vector in a spin rotation.}\BibitemShut {Stop}%
\bibitem [{not({\natexlab{i}})}]{note3}%
  \BibitemOpen
  \href@noop {} {} \bibinfo {note} {The same remark holds
  for the NPRG approach in the 1PI formalism. In this case, the action is
  defined by $S_k=S+\Delta S_k$ with a regulator $\Delta S_k[\varphi] = \half
  \sum_{\alpha,\alpha'} \varphi_\alpha R_{k,\alpha\alpha'} \varphi_{\alpha'}$
  quadratic in the field. The 1PI effective action is defined as $\Gamma_k^{\rm
  LT}[\phi]-\Delta S_k[\phi]$ with $\phi_\alpha=\mean{\varphi_\alpha}$ and
  $\Gamma_k^{\rm LT}[\phi]$ the Legendre transform of the free energy. The
  subtraction of $\Delta S_k[\phi]$ allows us to compensate (exactly at the
  mean-field level) the regulator term in the action and implies that
  $\Gamma_k$ (rather than $\Gamma_k^{\rm LT}$) has the meaning of a
  coarse-grained free energy. Correspondingly, the physical state of the system
  is determined by the minimum of $\Gamma_k$ (and not $\Gamma_k^{\rm
  LT}$).}\BibitemShut {Stop}%
\bibitem [{not({\natexlab{j}})}]{note8}%
  \BibitemOpen
  \href@noop {} {} \bibinfo {note} {The bosonic trace is
  defined in Appendix~\ref{app_bform} [Eq.~(\ref{Trdef})].}\BibitemShut {Stop}%
\bibitem [{not({\natexlab{k}})}]{note13}%
  \BibitemOpen
  \href@noop {} {} \bibinfo {note} {One can easily show that
  Eq.~(\ref{dphi1}) reproduces perturbation theory to order
  $\calO(U^2)$.}\BibitemShut {Stop}%
\bibitem [{not({\natexlab{l}})}]{note4}%
  \BibitemOpen
  \href@noop {} {} \bibinfo {note} {It is obvious that the
  ``noninteracting'' part of $\Gamma_k[\calG]$, namely $\half \tr \ln (-\calG)
  - \half \tr (\calG_k^{(0)-1}\calG-1)$, satisfies all symmetries of the
  action.}\BibitemShut {Stop}%
\bibitem [{not({\natexlab{m}})}]{note14}%
  \BibitemOpen
  \href@noop {} {} \bibinfo {note} {This conclusion is
  drawn, for instance, from the 1PI fermionic RG.}\BibitemShut {Stop}%
\bibitem [{not({\natexlab{n}})}]{note9}%
  \BibitemOpen
  \href@noop {} {} \bibinfo {note} {For a similar reason, it
  is difficult to compute the self-energy in the 1PI formalism (as it requires
  to retain the momentum and frequency dependence of the 1PI vertex
  $\bar\calY_k$); see, however, Ref.~\onlinecite{Giering12}.}\BibitemShut
  {Stop}%
\bibitem [{not({\natexlab{o}})}]{note5}%
  \BibitemOpen
  \href@noop {} {} \bibinfo {note} {As $k$ decreases, the
  self-energy correction due to the Goldstone modes becomes stronger and
  stronger thus suppressing the order parameter associated to the spontaneously
  broken symmetry. At a finite momentum scale $k$ (of the order of the inverse
  correlation length), the symmetry is restored and the Goldstone modes
  disappear.}\BibitemShut {Stop}%
\bibitem [{not({\natexlab{p}})}]{note16}%
  \BibitemOpen
  \href@noop {} {} \bibinfo {note} {The Wigner distribution
  function is defined by $n_{\p\sig}(\r,\tau)=\intr' e^{-i\p\cdot\r'}
  \mean{\psi^*_\sig(\r-\r'/2,\tau) \psi_\sig(\r+\r'/2,\tau)}$.}\BibitemShut
  {Stop}%
\end{thebibliography}

\end{document}